\documentclass[times,review]{elsarticle}

\usepackage{framed,multirow}
\usepackage{amssymb}
\usepackage{latexsym}

\usepackage[T1]{fontenc}
\usepackage[utf8]{inputenc}
\usepackage{amsmath,amsthm,amsfonts}
\usepackage{upgreek}
\usepackage[hidelinks]{hyperref}
\usepackage[caption=false]{subfig}
\usepackage[makeroom]{cancel}
\usepackage{mesh}
\usepackage[inline]{enumitem}
\usepackage[section]{placeins}

\usepackage{lineno}

\usepackage[markup=default, addedmarkup=bf, final]{changes}
\newcommand{\msout}[1]{\text{\sout{\ensuremath{#1}}}}
\newcommand{\stkout}[1]{\ifmmode\msout{#1}\else\sout{#1}\fi}
\setdeletedmarkup{\stkout{#1}}
\setauthormarkuptext{name}
\definechangesauthor[name={Authors},      color=orange]{A}
\definechangesauthor[name={Reviewer \#1}, color=blue]{R1}
\definechangesauthor[name={Reviewer \#2}, color=red]{R2}
\definechangesauthor[name={Reviewer \#3}, color=teal]{R3}
\definechangesauthor[name={Reviewer \#4}, color=magenta]{R4}

\newcommand{\Field}{\mathbb{K}}
\newcommand{\set}[1]{{#1}}                  
\newcommand{\nel}[1]{|\set{#1}|}            
\newcommand{\sfsize}[2][\Reals]{{#1}^{\nel{#2}}}          
\newcommand{\msize}[3][\Reals]{{#1}^{#2 \times #3}}       
\newcommand{\tsize}[3][\Reals]{\msize[#1]{\nel{#2}}{\nel{#3}}}  
\newcommand{\dscl}[2][s]{\mathsf{{#2}_{#1}}}      
\newcommand{\dvec}[2][s]{\vec{\mathsf{#2}}_\mathsf{{#1}}}
\newcommand{\Id}[1]{\mathbb{I}_{#1}}        
\newcommand{\Ic}[1]{\mathsf{I_{#1}}}        
\newcommand{\Iv}[1]{\mathsf{1}_{\set{#1}}}  
\newcommand{\Zv}[1]{\mathsf{0}_{\set{#1}}}  
\newcommand{\sfield}[2][]{\dscl[#1]{#2}}     
\newcommand{\vfield}[2][]{\dvec[#1]{#2}}    
\newcommand{\DOT}[3][]{\left ( #2, #3 \right )_\mathsf{#1}}
\newcommand{\T}{^\mathsf{T}}                
\newcommand{\strain}{S}                     
\newcommand{\rotation}{W}                   
\newcommand{\deviatoric}{\tau}              
\newcommand{\stress}{\sigma}                
\newcommand{\st}{\gamma}                    
\newcommand{\vel}{\vec{u}}                  
\newcommand{\A}{A}                          
\newcommand{\Sd}{\tilde{\A}}                
\newcommand{\topo}[1]{\mathsf{T_\set{#1}}}  
\newcommand{\metric}[1]{\mathsf{M_\set{#1}}}
\newcommand{\x}[1]{\vfield[#1]{x}}          
\newcommand{\ki}[1][]{\kappa_\set{#1}}      
\newcommand{\ks}{\sfield[f]{k}}             
\newcommand{\kc}{\sfield[c]{k}}             
\renewcommand{\ni}{\hat{\eta}_{i}}          
\newcommand{\nw}{\hat{n}_{wall}}            
\newcommand{\ns}{\sfield[f]{\uvec{n}}}      
\newcommand{\di}[1]{\sfield[#1]{r}}         
\newcommand{\ls}{\theta}                    
\newcommand{\lsv}[1]{\sfield[#1]{\uptheta}} 
\newcommand{\LS}[1][C]{\sfield[C]{\Theta}}  
\newcommand{\Nc}{\sfield{N_C}}              
\newcommand{\grad}{\nabla}                  
\renewcommand{\div}{\nabla \cdot}           
\newcommand{\adv}{\vel \cdot \nabla}        
\newcommand{\DIV}[1][]{\mathsf{D}_{\mathsf{#1}}}     
\newcommand{\GRAD}[1][]{\mathsf{G}_{\mathsf{#1}}}    
\newcommand{\LAP}[1][]{\mathsf{L}_{\mathsf{#1}}}     
\newcommand{\CONV}[1][]{\mathsf{C}(\uf)_{\mathsf{#1}}}
\newcommand{\U}[1][]{\mathsf{U}}            
\newcommand{\E}{{\Upsilon}}                 
\newcommand{\HR}{\Psi}                      
\newcommand{\sCF}{\Pi}                      
\newcommand{\aCF}{\Lambda}                  

\newcommand{\pc}{\sfield[c]{p}}             
\newcommand{\uc}{\sfield[c]{u}}             
\newcommand{\uf}{\sfield[f]{u}}             
\newcommand{\Df}{\Delta \dscl[F]{x}}        
\newcommand{\Le}{\mathsf{W_\set{E}}}        
\newcommand{\Sf}{\mathsf{\A_\set{F}}}       
\newcommand{\Vc}{\mathsf{V_\set{C}}}        
\newcommand{\K}[1][]{\mathsf{K_{#1}}}  
\newcommand{\Coeff}{\mu}                
\newcommand{\Ci}[1]{\Coeff_\set{#1}}        

\begin{document}


\begin{frontmatter}
\title{An energy-preserving level set method for multiphase flows}

\author[cttc]{N. Valle}
\ead{nicolas.valle@upc.edu}
\author[cttc]{F.X. Trias}
\ead{xavi@cttc.upc.edu}
\author[cttc]{J. Castro}{\corref{cor1}}
\ead{jesus@cttc.upc.edu}
\cortext[cor1]{Corresponding author}
\address[cttc]{Heat and Mass Transfer Technological Centre (CTTC), Universitat 
Politècnica de Catalunya - BarcelonaTech (UPC), ESEIAAT, Carrer Colom 11, 08222 
Terrassa (Barcelona)}


\begin{abstract}
  The computation of multiphase flows presents a subtle energetic equilibrium 
  between potential (i.e., surface) and kinetic energies. The use of traditional 
  interface-capturing schemes provides no control over such a dynamic balance.  
  In the spirit of the \replaced[id=R1]{well-known}{well-know} 
  symmetry-preserving and mimetic schemes, whose physics-compatible 
  discretizations rely upon preserving the underlying mathematical structures of 
  the space, we identify the corresponding structure and propose a new 
  discretization strategy for curvature. The new scheme ensures conservation of 
  mechanical energy (i.e., surface plus kinetic) up to temporal integration.  
  Inviscid numerical simulations are performed to show the robustness of such a 
  method.
\end{abstract}

\begin{keyword}
Multiphase Flow \sep Symmetry-preserving \sep Mimetic \sep Conservative Level 
Set \sep Energy-preserving
\end{keyword}

\end{frontmatter}

\newcommand{\ifStaticBubble}{\iftrue}
\newcommand{\ifOscillatingEllipse}{\iftrue}
\newcommand{\ifCapillaryWave}{\iftrue}

\section{Introduction}
\label{sec:Introduction}

Multiphase flows are ubiquitous in industrial applications. They are present in 
a rich variety of physical phenomena such as vaporization \cite{Tanguy2007}, 
atomization \cite{Desjardins2008}, electrohydrodynamics \cite{VanPoppel2010} or 
boiling \cite{Tanguy2014}, among others \cite{Lepilliez2016, Gutierrez2017}.

The use of interface-capturing schemes is widespread for the computation of 
multiphase flows due to its computational efficiency. The
Volume-Of-Fluid (VOF) by Hirt and Nichols \cite{Hirt1981}, the Level Set method 
developed by Osher and Sethian \cite{Osher1988} and most recently phase field 
methods, introduced by Anderson et al. \cite{Anderson1997}, are the most popular
interface capturing schemes for multiphase flows. An overview of these can be 
found in \cite{Mirjalili2017} and references therein. Despite the pros and cons 
that each method presents, we made our development concrete on the Conservative 
Level Set (CLS) initially developed by Olsson and Kreiss \cite{Olsson2005} and 
Olsson et al. \cite{Olsson2007} due to its good conservation properties, 
curvature accuracy and ease of handling topological changes.
This was extended to unstructured collocated meshes in \cite{Balcazar2014b}.

Of particular interest are the incompressible Navier-Stokes equations,
\begin{align}
  \rho \left( \frac{\partial \vel}{\partial t}
  + (\vel \cdot \grad) \vel \right) &= \div \stress & \div \vel &= 0
\label{eqn:NS}
\end{align}
where the stress tensor $\stress$ is composed of the hydrostatic and the 
deviatoric ones ($\stress = -p\Id{} + \deviatoric$). In turn, $\deviatoric$ is 
defined by Stokes constitutive equation $\deviatoric = 2 \mu \strain$, 
\replaced[id=A]{while}{where} the strain tensor is given by $\strain = 1/2 
\left( \nabla \vel + (\nabla \vel)^T \right)$.

The proper solution of equations (\ref{eqn:NS}) requires an appropriate 
decoupling of pressure and velocity. In this regard, the Fractional Step Method 
(FSM) \cite{Chorin1968} is an excellent tool which properly enforces the 
incompressibility constraint. However, the FSM results in a Poisson equation 
which needs to be solved, which takes most of the computational time in a 
typical simulation.

The construction of discrete differential operators in the seminal work of 
Verstappen and Veldman \cite{Verstappen1997, Verstappen2003} aims at preserving 
physical quantities of interest, namely momentum and kinetic energy, by 
preserving several mathematical properties at the discrete level. This merges 
with the conception of mimetic finite difference methods \cite{Lipnikov2014}, 
where the discretization is performed to satisfy the inherent mathematical 
structure of the space, naturally producing a physics-compatible discretization.  
The present work is motivated by such an appealing idea. This has been named 
mimetic \cite{Lipnikov2014} or discrete vector calculus \cite{Robidoux2011}, 
among others \cite{Tonti2014,Perot2011}. Mimetic methods delve into the 
construction of discrete differential operators by producing discrete 
counterparts of more fundamental mathematical concepts, making extensive use of 
exterior calculus.  This approach results in the algebraic concatenation of 
elementary operators, namely matrices and vectors.  Such an approach can be seen 
as the mathematical dual of the physics-motivated work on symmetry-preserving 
schemes and provides with a different point of view which 
\replaced[id=R1]{fortifies}{fortify} the analysis of this family of 
methods\replaced[id=A]{, which}{. These ideas} have been used in both 
academic~\cite{Trias2015} and
industrial problems~\cite{Paniagua2014,Giraldez2016}, among others.  However, to 
our knowledge, there is no a straightforward extension of these ideas into the 
multiphase flow community yet.

While Direct Numerical Simulation (DNS) of single-phase flows has reached 
substantial maturity, multiphase flows lag behind due to its increased 
complexity, namely due to two main issues:
\begin{enumerate*}[label={\roman*)}, font=\itshape]
  \item surface tension
  and \item differences in physical properties
\end{enumerate*}.
The former results in a dynamic equilibrium between kinetic and potential 
energies, which are exchanged through the capillary term. Indeed this is the 
reason why symmetry-preserving discretizations \cite{Verstappen1997, 
Verstappen2003}, despite conserving flawlessly kinetic (and thus total) energy 
in single-phase flows, do not suffice to preserve mechanical energy in 
multiphase flows, as this transfer needs to be taken into account explicitly.
The later poses challenges regarding how interpolations need to be done without 
breaking physical laws. In the framework of VOF, Fuster \cite{Fuster2013} 
proposed a discretization that preserves the (skew-)symmetries of the momentum 
equation, preserving kinetic energy up to surface tension, which is regarded as 
an energy source.  However, as far as surface tension is not included into the 
analysis, this is a necessary, but not a sufficient condition for preserving 
mechanical energy.  Regarding the viscous term, the work of Sussman et al.  
\cite{Sussman2007} provided with a conservative discretization. The interested 
reader is referred to \cite{Lalanne2015} and references therein for a comparison 
between different discretization strategies for the viscous term. Despite the 
impressive progress done so far, none of the above have included surface 
tension, and thus potential energy, in the analysis of conservation of energy.  
It is well-known, however, that the imbalances in the surface tension term may 
lead to spurious currents and, eventually, unstable solutions 
\cite{Magnini2016}. In the framework of phase field methods, the impact of 
surface tension on the energy balance has been included in the works of Jacqmin 
\cite{Jacqmin1999} for the Cahn-Hilliard equation, and Jamet et al.  
\cite{Jamet2002} and Jamet and Misbah \cite{Jamet2008} for the Allen-Cahn 
formulation. This paper aims to dig into a discretization including surface 
tension which\added[id=R1]{, without recompression,} preserves mechanical energy
for level set schemes.

The rest of the paper is arranged as follows: in Section \ref{sec:Topology} a 
glimpse of algebraic topology is provided. This sets the foundations to review 
the well-known symmetry-preserving discretization for single-phase flows in 
Section \ref{sec:DiffModel_single} and, inspired by this, develop a new 
energy-preserving scheme for multiphase flows in Section 
\ref{sec:DiffModel_multi}. Comparative results between current techniques and 
the newly developed methods are presented in Section \ref{sec:Results}.  
Finally, conclusions and future insights are outlined in Section 
\ref{sec:Conclusions}.

\section{Topological model}
\label{sec:Topology}

Any numerical approach requires a finite-dimensional representation of the 
spaces under consideration.
This implies a discrete representation of the domains involved in the setup of 
the problem.
Single-phase flows fit well into a fixed frame, typically discretized on a fixed 
grid. On the other hand, multiphase flows require to account for a moving 
interface which splits the domain at question into two regions. This interface 
needs to be properly discretized in order to preserve several inherent 
topological properties. The way this is accomplished has led to a diversity of 
multiphase methods \cite{Mirjalili2017}.

\subsection{Mesh}
\label{sec:Model_Topology_Mesh}

\begin{figure}[h]
 \centering
 \raisebox{-0.5\height}{\includegraphics{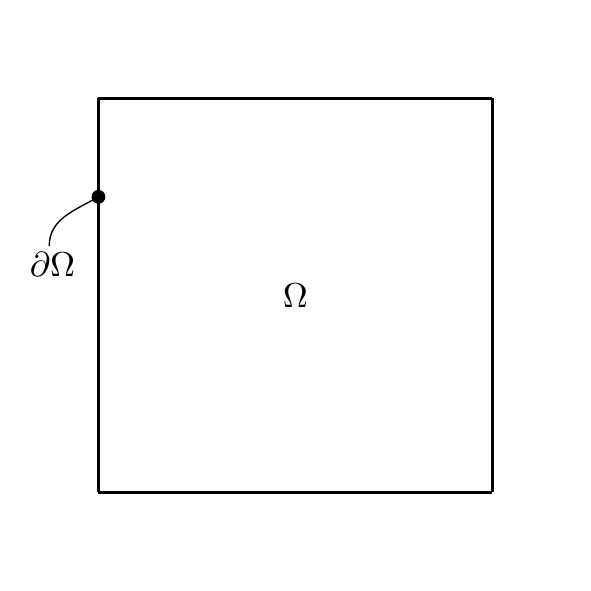}}
 \raisebox{-0.5\height}{\includegraphics{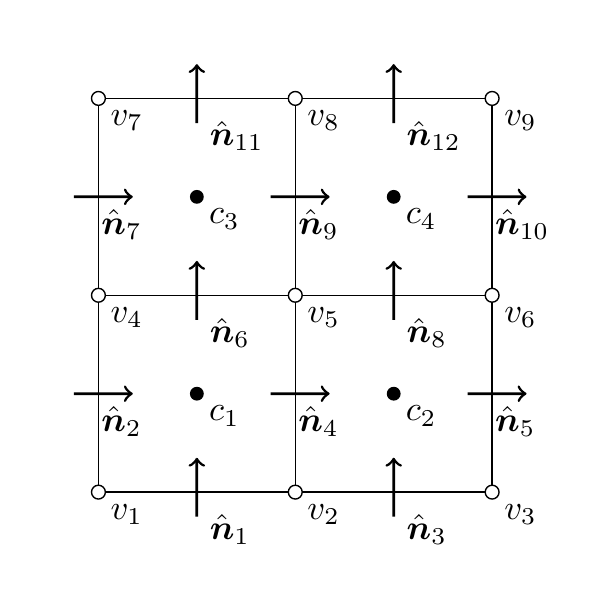}}
 \caption{Left: Domain $\Omega$ and its boundary $\partial\Omega$. Right: Mesh 
   $\set{M}$.  $c_i$ corresponds with the $ith$ cell, $\uvec n_i$ corresponds 
 with the normal vector to the $jth$ face (i.e., $f_j$) and $v_k$ corresponds 
 with the $kth$ vertex. Its incidence matrix is stated in equation 
 (\ref{eqn:Tfc_Example_struct}).}
 \label{fig:model_mesh_struct}
\end{figure}

Let $\Omega$ be the working domain bounded by $\partial \Omega$ and assume that 
$\set{M}$ is a partition of $\Omega$ into a non-overlapping mesh. An 
illustrative example is given in Figure \ref{fig:model_mesh_struct}. Although we 
stick to structured meshes for computational simplicity, the formulation 
presented here is independent of the mesh structure and thus can be extended 
into unstructured meshes.
Incidence matrices are used to account for the connectivity within geometric 
entities. An example for \added[id=R1]{$\topo{FC}$, the incidence matrix 
relating faces with cells according to the orientation of the mesh given in} 
Figure \ref{fig:model_mesh_struct} is shown next
\begin{equation}
 \label{eqn:Tfc_Example_struct}
 \topo{FC} = \bordermatrix{
     & f_1& f_2& f_3& f_4& f_5& f_6& f_7& f_8& f_9& f_{10}& f_{11}& f_{12}\cr
     c_1 & -1 & \replaced[id=R1]{-1}{+1} &  0 & \replaced[id=R1]{+1}{-1} &  0 & 
 +1 &  
 0 &  0 &  0 &  0    &  
 0    &  
 0    \cr
 c_2 &  0 &  0 & -1 & \replaced[id=R1]{-1}{+1} & \replaced[id=R1]{+1}{-1} &  
 0 &  
 0 & +1 & -1 &  0    &  
 0    &  0    \cr
 c_3 &  0 &  0 &  0 &  0 &  0 & -1 & \replaced[id=R1]{-1}{+1} &  0 &  
 \replaced[id=R1]{+1}{0} &  0    & +1    &  0    \cr
 c_4 &  0 &  0 &  0 &  0 &  0 &  0 &  0 & -1 & \replaced[id=R1]{-1}{+1} & 
 \replaced[id=R1]{+1}{-1}    &  
  0    & +1    \cr
  }
\end{equation}
They replace usual neighboring relations such as $\phi_c = \sum_{f\in c} \phi_f$ 
for the sum of face values related to cell $c$\replaced[id=R1]{. In addition, 
its transpose provides with an explicit form for}{ or} $\Delta_f \phi =
\phi_+ - \phi_-$ for the difference across face $f$, among others.  Basic 
geometric properties such as edge lengths ($\Le$), face surfaces ($\Sf$) and 
cell volumes ($\Vc$) are arranged as diagonal matrices.
This matrix perspective presents several advantages:
\begin{enumerate*}[label={\roman*)}, font=\itshape]
  \item mesh independence
    ,\item computational simplicity
  and \item readily accessible algebraic analysis
\end{enumerate*}.
While we restrain ourselves from digging into the first two, the later is useful 
both for reviewing the classical symmetry-preserving scheme and the development 
of the novel technique described here. Hereafter, lowercase letters correspond 
with vectors, whose subscript indicates the geometric entity to which they are 
linked (e.g., $\sfield[c]{p}$ corresponds to pressure located at cells).  
Capital letters correspond with matrices, whose subscript(s) identify rows and 
(if different) columns (e.g. $\topo{FC}$ is the face-to-cell incidence matrix).

\subsection{Interface}
Interfaces imply a moving topology along the working domain, which implies a 
Lagrangian frame of reference. Interface tracking schemes track such a frame 
explicitly, at the expenses of numerical complexity \cite{Tryggvason2001}. On 
the other hand, interface capturing schemes preserve a fully Eulerian approach, 
by mapping quantities expressed in the Lagrangian frame back into the Eulerian 
one \cite{Hirt1981,Sussman1994,Chen1997}. This results in a simpler 
implementation of the interface at the cost of an implicit representation. At 
this point we split the presentation between the techniques used to actually 
capture the evolution of the interface and the ones used to obtain explicit 
geometric information out of the implicit form.

\subsubsection{Interface Capturing}
\label{sec:Model_Topology_Capturing}

\begin{figure}[h]
  \centering
  \includegraphics[width=8cm]{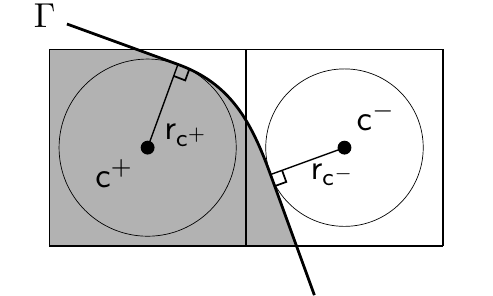}
  \caption{Distances $\di{c^\pm}$ are defined as the shorter distances of the 
interface to the cell. These are then normal to the interface and correspond 
with the minimum radius of the tangent sphere.}
  \label{fig:model_interface_capture}
\end{figure}

Let's assume now that the domain $\Omega$ presents an interface at $\Gamma$, 
which splits $\Omega$ into $\Omega^+$ and $\Omega^-$. We note that the volume of 
a single phase $\Omega^+$ can be defined as
\begin{equation}
  \int_{\Omega^+} dV = \int_{\Omega} H(r) dV
  \label{eqn:VolumeIntegral}
\end{equation}
where $r$ corresponds with the signed shorter distance from an arbitrary point 
to the interface, as can be seen in Figure \ref{fig:model_interface_capture}, 
while $H(r)$ is its corresponding Heaviside step function, which is valued $1$ 
at phase $\Omega^+$ and $0$ otherwise.
Note that this function is the key to map a Lagrangian frame ($\Omega^+$) back 
into an Eulerian one ($\Omega$).
Specific tracking of such a quantity is the basis of the Volume of Fluid (VOF) 
method \cite{Hirt1981}, which yields to the concept of volume fraction or, more 
generally, marker function. Despite being formally neat, the implementation of 
specific convection schemes is required, eventually requiring full geometric 
reconstruction, resulting in an intricate implementation. A different approach 
is to capture the interface with a CLS \cite{Olsson2005, Olsson2007}. This 
captures the interface as the isosurface of a continuous and smooth function 
$\ls$. The level set marker function, $\ls$, results in a smoothed Heaviside 
step function that preserves $\int_\Omega \ls dV = \int_\Omega H(r) dV$. It is 
constructed as the convolution of the distance function $r$ as follows
\begin{equation}
  \label{eqn:LS-HyperbolicConvolution}
  H(r) \approx \ls(r) = \frac{1}{2} \left( tanh \left( \frac{r}{2\epsilon} 
\right) + 1 \right)
\end{equation}
where $\epsilon$ corresponds with a smoothing factor. Note that $\ls(r) \to 
H(r)$ as $\epsilon \to 0$. Further details can be found in \cite{Balcazar2014b, 
Wacawczyk2015}.

By imposing the conservation of the marker function, we can advect such a marker 
in an incompressible flow as \cite{Olsson2005, Olsson2007, Balcazar2014b}:
\begin{equation}
  \frac{\partial \ls}{\partial t} + (\vel \cdot \nabla) \ls = 0
  \label{eqn:LS-advection}
\end{equation}
where particular advection schemes and 
\replaced[id=R1]{recompression}{re-initialization} stages \replaced[id=R3]{can 
be added in order to obtain a sharper}{may be needed in order to maintain a 
sharp enough} profile.  The interested reader is referred to 
\cite{Olsson2005,Balcazar2014b} and references therein.

%
%
%
%

\subsubsection{Interface Reconstruction}
\label{sec:Model_Topology_Reconstruction}

Surface reconstruction may start by defining the interface normal. It is 
computed as \cite{Olsson2005}
\begin{equation}
  \ni = \frac{\nabla\ls}{|\nabla\ls|}
  \label{eqn:LS-normal}
\end{equation}

Which implies, by definition, that the gradient of the marker function $\ls$ is 
parallel to the normal. On the other hand, curvature is computed as
\begin{equation}
  \kappa = - \div \ni
  \label{eqn:LS-curvature}
\end{equation}

Now, the surface area of $\Gamma$ can be computed in any of the following forms
\begin{equation}
  \A = \int_{\Gamma} d\A  = \int_{\Omega} \delta(r) dV
                          = \int_{\Omega} \nabla H(r) \cdot \ni dV
\label{eqn:SurfaceIntegral}
\end{equation}
\replaced[id=R1]{where $\delta(r)$ is Dirac's delta function, which formally is 
the distributional derivative of the Heaviside step function. This}{which} is
the basis of the celebrated continuum surface force of Brackbill et al.  
\cite{Brackbill1992} for surface tension and, in general, of other smoothed 
interface methods.  Regardless of the reconstruction method of choice, surface 
needs to satisfy first variation of area formula, which relates surface and 
volume variations through curvature and velocity as
\begin{equation}
  \frac{d}{dt}\int_\Gamma d\A = - \int_\Gamma \kappa \vel \cdot \ni d\A
  \label{eqn:dSdt}
\end{equation}

This is a
\replaced[id=R3]{fundamental identity, and the ultimate responsible of the
correct conversion between kinetic and surface energy, as it will be shown in 
Section \ref{sec:DiffModel_multi_conservation}}{crucial identity that must be 
satisfied at all times for closed surfaces}.
\deleted[id=R1]{The extension to these cases is out of the scope of this work.}
\added[id=R1]{A detailed proof of this can be found in chapter 8.4 of  
  Frankel~\cite{Frankel2012}.}
\added[id=R1]{As an illustrative example, let us
consider the surface variation of a spherical surface.  If we analyze how $\A = 
4\pi r^2$ evolves under the action of the normal velocity, $\dot{r}$, we obtain 
that $\frac{d\A}{dt} =   8\pi r \dot{r}$, which can be rearranged as 
$\frac{d\A}{dt} = \frac{2}{r} \A \dot{r}$, where we identify $\kappa = 
\frac{2}{r}$, the mean curvature of sphere}. We are now going to prove that the 
use of a smooth marker function as in equation 
(\ref{eqn:LS-HyperbolicConvolution}) leads to a consistent modeling of the 
interface by reconstructing surface area with its smooth counterpart $\Sd$ as
\begin{equation}
  \A\overset{(\ref{eqn:SurfaceIntegral})}{=}
    \int_\Omega \nabla H(r) \cdot \ni dV
    \overset{(\ref{eqn:LS-HyperbolicConvolution})}{\approx}
    \int_\Omega \grad \ls \cdot \ni dV
    \overset{(\ref{eqn:LS-normal})}{=}
    \int_\Omega |\grad \ls| dV
    =
    \Sd
  \label{eqn:LS-SurfaceReconstruction}
\end{equation}

We now show that equation (\ref{eqn:LS-SurfaceReconstruction}) is a compatible 
approximation of $\A$. In particular, we prove that equation (\ref{eqn:dSdt}) is
still valid when we replace $\A$ by $\Sd$\added[id=R4]{, which is defined over 
the volumes and thus much more convenient to compute}. First, as a preliminary 
stage, we take the gradient of the transport equation (\ref{eqn:LS-advection}) 
in the pursue of a relation between the marker function and the smoothed surface
\begin{equation}
  \frac{\partial \nabla \ls}{\partial t} +
  \nabla\left( (\adv) \ls \right) = 0
  \label{eqn:LS-GdHdt}
\end{equation}

\added[id=R4]{
  Finally, before moving on to the announced proof, let us introduce the inner 
  product notation $\DOT{\cdot}{\cdot}$, which simplifies bi-linear integrals as
  $\int fg d\set{S} = \DOT{f}{g}$. In addition, concepts such as orthogonality, 
  duality or (skew-)symmetry are naturally expressed in this framework.  Further
  details can be found in \ref{sec:InnerProducts}.
}
\replaced[id=R4]{
  With this in mind, we can proceed}
  {This can used}
to approximate the left hand side of equation (\ref{eqn:dSdt}) via equation 
(\ref{eqn:LS-SurfaceReconstruction}), to yield the following
\begin{equation}
  \begin{split}
    \frac{d}{dt}\int_\Gamma d\A
  \overset{(\ref{eqn:LS-SurfaceReconstruction})}{\approx} \frac{d\Sd}{dt}
  &= \frac{d}{dt} \DOT{\nabla \ls}{\ni}
   = \DOT{\frac{d\nabla \ls}{dt}}{\ni}
   + \DOT{\nabla \ls}{\frac{d\ni}{dt}} \\
  &= \DOT{\frac{\partial \nabla \ls}{\partial t}}{\ni}
   + \cancel{\DOT{(\adv) \nabla \ls}{\ni}}
   + \DOT{\nabla \ls}{\frac{\partial \ni}{\partial t}}
   + \cancel{\DOT{\nabla \ls}{(\adv) \ni}} \\
  &= \DOT{\frac{\partial \nabla \ls}{\partial t}}{\ni}
   + \cancel{\DOT{\nabla \ls}{\frac{\partial \ni}{\partial t}}}
   \overset{(\ref{eqn:LS-GdHdt})}{=} - \DOT{\nabla \left( \left( \adv \right) 
  \ls \right)}{\ni} \\
  &= \DOT{\div \ni}{\vel\cdot \nabla \ls}
  \qed
  \end{split}
  \label{eqn:dSdt-lhs-smoothed}
\end{equation}
\replaced[id=R4]{where we exploit the skew-symmetry of the convective operator 
  in the second row of eq.~(\ref{eqn:dSdt-lhs-smoothed}), benefit form 
$\frac{\partial \ni}{\partial t} \perp \nabla \ls$ in the third one and the 
duality between gradient and divergence in the last one.}
{where we benefit from $\frac{\partial \ni}{\partial t} \perp \nabla \ls$ by 
  equation (\ref{eqn:LS-normal}) as well as the skew-symmetry of the convective 
  operator $\left ( (\vel \cdot \nabla) = - (\vel \cdot \nabla)^*\right )$ for 
  incompressible flows (i.e., $\div \vel = 0$) together with $\ni \parallel 
\grad\ls$.}

\deleted[id=A]{
Equation (\ref{eqn:dSdt-lhs-smoothed}) can be rewritten with the help of a 
re-scaled gradient}
\deleted[id=A]{
to yield}
\deleted[id=A]{
where we can split $\nabla \vel$ into symmetric ($\strain$, strain) and 
skew-symmetric components ($\rotation$, rotation) as $\nabla \vel = \strain + 
\rotation$. We note that the later has a null contribution when it is right- and
left- multiplied by the same vector, so we obtain}
\deleted[id=A]{
which implies that only strain contributes to modify $\Sd$.
From this we can identify the stretching term of the surface, which, by virtue 
of equation (\ref{eqn:LS-normal}), it is nothing but the normal component of the
strain tensor, $\strain$, scaled by $\Sd$.}  

Regarding the approximation of the right hand side of equation (\ref{eqn:dSdt}),
we can proceed by including equation (\ref{eqn:LS-curvature})\replaced[id=R4]{ 
and then using equation (\ref{eqn:LS-SurfaceReconstruction}) to move from 
surface to volume integrals}{via equation (\ref{eqn:LS-SurfaceReconstruction})} 
as
\begin{equation}
    - \int_\Gamma \kappa \vel \cdot \ni
    \replaced[id=R4]{d\A}{d\Sd}
    \overset{\added[id=R4]{(\ref{eqn:LS-curvature})}{}}{=}
    \int_\Gamma (\div \ni) \vel \cdot \ni
    \replaced[id=R4]{d\A}{d\Sd}
    \overset{\added[id=R4]{(\ref{eqn:LS-SurfaceReconstruction})}{}}{\approx}
      \DOT{\div \ni}{\vel\cdot \nabla \ls}
    \label{eqn:dSdt-rhs-smoothed}
\end{equation}

\replaced[id=A]{We}{Finally, by exploiting duality between gradient and 
divergence ($\div = -\grad^*$) and, again, the skew-symmetry of the convective 
operator for incompressible flows together with the fact that $\ni \parallel 
\grad \ls$ to equations (\ref{eqn:dSdt-lhs-smoothed}) and 
(\ref{eqn:dSdt-rhs-smoothed}), we} finally obtain:
\begin{equation}
  \frac{d}{dt}\int_\Gamma d\Sd
  = \DOT{\left( \adv  \right)\ls}{\div \ni}
  = - \int_\Gamma \kappa \vel \cdot \ni d\A
  \label{eqn:dSdt-identity}
\end{equation}
We conclude that, from a continuum point of view, the use of a marker function 
together with the surface reconstruction strategy stated in equation 
(\ref{eqn:LS-SurfaceReconstruction}) results in a consistent capture of both the 
interface and its geometric features, namely the first variation of area 
equation (\ref{eqn:dSdt}). Notice that this analysis is not exclusive to level 
sets, but rather extensible to other interface capturing schemes as far as the 
surface can be cast into a potential form as in equation 
(\ref{eqn:LS-SurfaceReconstruction}).

\section{Symmetry-preserving discretization of single-phase flows}
\label{sec:DiffModel_single}
In an incompressible flow, in the absence of external forces, the net balance of 
mechanical energy is due to the viscous term of the Navier-Stokes equation 
solely. This is a relevant property for the simulation of turbulent flows, 
particularly for the computation of DNS.  In this section, the well-known finite 
volume, staggered, symmetry-preserving discretization of Verstappen and Veldman 
\cite{Verstappen1997,Verstappen2003} is briefly reviewed. This 
\replaced[id=R1]{sets}{set} the ground of the newly developed energy-preserving 
scheme presented in section \ref{sec:DiffModel_multi}. Assuming constant 
physical properties, Navier-Stokes equations (\ref{eqn:NS}) can be rearranged to 
yield
\begin{align}
  \rho\left( \frac{\partial \vel}{\partial t} + \left ( \vel\cdot\nabla \right ) 
        \vel \right) &= - \nabla p + \mu \nabla^2 \vel
       & \nabla \cdot \vel &= 0
\label{eqn:NS-single}
\end{align}
which is the most common form of the Navier-Stokes equations for incompressible 
\replaced[id=R1]{single-phase}{multiphase} flows.

\subsection{Energy conservation}
\label{sec:DiffModel_single_conservation}
The evolution of kinetic energy, $E_k = \DOT{\vel}{\rho\vel}$, in a single-phase 
flow is obtained by taking the inner product of $\vel$ and equation 
(\ref{eqn:NS-single})\added[id=R4]{, which, in the absence of external forces 
and without contributions from the boundaries, yields:}
\begin{equation}
  \frac{d E_k}{dt} =  - \rho (\vel,(\vel \cdot \nabla) \vel)
                      - (\vel, \nabla p)
                      + \mu (\vel, \nabla^2 \vel) =
                      - \mu \left\| \nabla \vel\right\|^2 \leq 0
  \label{eqn:dEkdt-single-continuum}
\end{equation}

Due to the skew-symmetry of the convective operator (i.e., $(\vel \cdot \nabla) 
= -(\vel \cdot \nabla)^*$), the contribution of this term to kinetic energy is 
null. Duality of the gradient operator with divergence (i.e., $\nabla^* = - 
\nabla \cdot$) together with the incompressible constrain of the velocity 
($\nabla \cdot \vel = 0$) results in a null contribution of the pressure term to 
kinetic energy \cite{Verstappen2003}.  Finally, by exploiting again the duality 
between gradient and divergence in the viscous term, this results in a 
negative-definite operator, $\mu(\vel, \nabla^2 \vel) = - \mu(\nabla \vel , 
\nabla \vel) = - \mu \left\| \nabla \vel\right\|^2$, which, as expected, dumps 
kinetic energy.

\subsection{Symmetry-preserving discretization}
\label{sec:DiffModel_single_SP}
We are now going to review the well-know symmetry-preserving, second-order, 
staggered, finite volume discretization introduced by Verstappen and Veldman in 
\cite{Verstappen1997}, which was subsequently extended to fourth order in 
\cite{Verstappen2003}, from the algebraic perspective by means of the tools 
introduced in Section~\ref{sec:Model_Topology_Mesh}. This lays the foundation of 
the newly developments introduced in Section~\ref{sec:DiffModel_multi}.
The discretization in a staggered grid starts by defining the discrete 
divergence operator, $\DIV$, directly from the Gauss-Ostrogradsky theorem
\begin{equation}
  \label{eqn:DiscreteDivergence}
  \int_\Omega \nabla \cdot \vec{u} dV =
  \int_{\partial\Omega} \vec{u} \cdot \vec{n} dS
  \approx  -\topo{FC}\Sf \uf =
  \metric{C} \DIV \uf,
\end{equation}
where $\metric{C} \in \tsize{C}{C}$ stands for the metric of the cells, which is 
a diagonal matrix containing cells volume. $\topo{FC} \in \tsize{C}{F}$ takes 
care of the appropriate sum of fluxes over the faces and $\Sf \in \tsize{F}{F}$ 
is the diagonal matrix containing the surface of all faces.  Finally $\uf$ 
stands for the staggered velocities. We can rearrange equation 
(\ref{eqn:DiscreteDivergence}) to yield
\begin{equation}
  \DIV = -\metric{C}^\mathsf{-1} \topo{FC}\Sf
  \label{eqn:DiscreteDivergenceExpanded},
\end{equation}
this leads to $\DIV \in \tsize{C}{F}$, as expected for a staggered grid 
arrangement.
Next, the discrete gradient operator, $\GRAD$, is constructed to preserve 
duality \begin{equation}
  \label{eqn:SymmPresGradientCondition}
  \DOT[F]{\uf}{\GRAD \pc} = - \DOT[C]{\DIV\uf}{\pc},
\end{equation}
where
$\DOT[C]{\sfield[c]{a}}{\sfield[c]{b}} = \sfield[c]{a^T} \metric{C} 
\sfield[c]{b}$
stands for the weighted inner product in the cell space, $\sfield{C}$. It can be 
defined conversely for the face space, $\sfield{F}$.
Further details on inner products can be found in \ref{sec:InnerProducts}.

In the context of Navier-Stokes equations, preserving this duality at the 
discrete level results into a null contribution of the pressure term to kinetic 
energy \cite{Verstappen2003}
\begin{equation}
  \label{eqn:DiscreteGradient}
  \GRAD = -\metric{F}^\mathsf{-1} \DIV\T \metric{C}
\end{equation}
where $\metric{F} \in \tsize{F}{F}$ corresponds to the metric of the face space, 
and thus the definition of such a metric induces the proper construction of 
$\GRAD$. This is nothing but the definition of the staggered control volume.  
Notice that $\GRAD \in \tsize{F}{C}$. Again, this 
\replaced[id=R1]{locates}{locate} gradients at faces, as expected for a
staggered discretization. $\metric{F}$ is defined as
\begin{equation}
  \label{eqn:Metric-Face}
  \metric{F} = \Df\Sf
\end{equation}

Note that $\Df \in \tsize{F}{F}$ is the diagonal arrangement of the distance 
between cell centers across the face, while $\Sf \in \tsize{F}{F}$ is also 
diagonal and contains face surfaces.  The final form of $\GRAD$ results in
\begin{equation}
  \label{eqn:DiscreteGradientExpanded}
  \GRAD = (\Df)^\mathsf{-1} \topo{CF}
\end{equation}
where the standard second-order approximation of the gradient arises naturally 
from the definition of the staggered control volume (i.e., one induces the 
other).
%
%

By concatenation, the discretization of the scalar Laplacian operator, $\LAP$, 
the essential element of the FSM, can be defined as follows
 \begin{equation}
   \label{eqn:DiscreteLaplacian}
   \int_\Omega \nabla^2 p dV \approx \metric{C} \LAP \pc = \metric{C} \DIV \GRAD 
     \pc
 \end{equation}
As expected, $\LAP \in \tsize{C}{C}$. We can expand the final integrated form of 
the discrete Laplacian as
\begin{equation}
  \metric{C} \LAP = - \topo{FC} \Sf (\Df)^\mathsf{-1} \topo{CF}
  \label{eqn:DiscreteLaplacianExpanded}
\end{equation}
%
%
%
%
where such a discretization results in a negative-definite operator since $\Sf$ 
and $\Df$ are positive-definite, and $\topo{FC} = \topo{CF}\T$.  This 
is the ultimate responsible of the diffusive character of viscosity in the 
context of Navier-Stokes equations. 

Finally, the convective term can proceed as in Hicken et al. \cite{Hicken2005a} 
in order to construct a skew-symmetric discretization. Even when dedicated 
convective operators may be constructed for Cartesian meshes, this provides with 
a more flexible approach. The idea is to construct proper face-to-cell and 
cell-to-face shift operators in order to exploit the collocated convective 
operator as
\begin{equation}
  \CONV[F] = \Gamma_{f \to c} \left( \Ic{d} \otimes \CONV[C] \right) \Gamma_{c 
  \to f}
  \label{eqn:DiscreteConv}
\end{equation}
which guarantees skew-symmetry as far as the vector-valued shift operators are 
transpose (i.e., $\Gamma_{f \to c}=\Gamma_{c \to f}\T$), as it is the 
case.  Further details can be found in \cite{Verstappen1997, Verstappen2003, 
Hicken2005a, Trias2014} and references therein.

\subsection{Analysis}
\label{sec:DiffModel_single_analysis}
By preserving (skew-)symmetries of the operators, as it was described above, the 
conservation of kinetic energy is guaranteed in the semi-discrete setup (i.e., 
up to temporal integration \cite{Capuano2017a}), mimicking then the continuous 
behavior of the system.  In particular, the semi-discretized energy balance 
equation reads
\begin{equation}
  \frac{d\sfield{E_k}}{dt} =  -\DOT[F]{\uf}{\CONV[F]\uf}
                              -\DOT[F]{\uf}{\GRAD \pc}
                              +\mu\DOT[F]{\uf}{\LAP[F]\uf}
                              \leq 0,
  \label{eqn:dEkdt-single-discrete}
\end{equation}
which is the discrete counterpart of equation 
(\ref{eqn:dEkdt-single-continuum}). As expected, the only term contributing to 
kinetic energy is the viscous one, i.e., $\mu\DOT[F]{\uf}{\LAP[F]\uf}$, where 
$\LAP[F]$ is the standard, negative-definite, staggered diffusive operator  
\cite{Verstappen2003}. Note that this holds thanks to the specific construction 
of the operators involved and if the incompressibility constrain of velocity is 
satisfied at the discrete level as well (i.e., $\DIV \uf = \Zv{c}$).

\section{Energy-preserving discretization of multiphase flows}
\label{sec:DiffModel_multi}
Multiphase flows present discontinuities at the interface due to the difference 
of physical properties and the existence of interfacial phenomena, namely, 
surface tension. This section develops, on top of the symmetry-preserving scheme 
reviewed in the previous section, a novel energy-preserving scheme for the 
discretization of curvature. Curvature plays a key role in the development of 
discontinuities, $[\cdot]$, across the interface as
\begin{equation}
  \left[ \stress \right]\ni = -\st \ki \ni
  \label{eqn:NS-multi-jumps}
\end{equation}
where $\st$ states for the surface tension coefficient, which we assume 
constant. This configures the resulting surface tension force, which acts at the 
interface by imposing a jump condition into the stress tensor which ``pulls'' 
the interface towards a lower free energy state.  The original governing 
equations (\ref{eqn:NS}) can then be reformulated as
  \begin{align}
    \rho \left( \frac{\partial \vel}{\partial t} + (\vel \cdot \nabla) \vel 
    \right) &=      - \grad p + \div \deviatoric
    & \div \vel &= 0 \label{eqn:NS-multiphase}\\
    \left[p\right] &= \ni^T\left[\deviatoric\right]\ni - \st \ki
    \label{eqn:NS-multiphase-jump}
  \end{align}
where $\stress = - p\Id{} + \deviatoric $ is split into hydrostatic and 
deviatoric.  The discussion about the discretization of $\deviatoric$ is out of 
the scope of this work, so the interested reader is refereed to Lalanne et 
al.~\cite{Lalanne2015} for a thoughtful discussion on this topic. At this point, 
it is assumed that $\deviatoric$ presents a prescribed discontinuity at the 
interface.

\subsection{Energy conservation}
\label{sec:DiffModel_multi_conservation}
Regardless of viscous dissipation, incompressible, multiphase flows, do not 
preserve kinetic energy. Instead, surface ($E_p=\st\A$) and kinetic 
($E_k=\DOT{\vel}{\rho\vel}$) energy are exchanged through the pressure 
term~\cite{Jacqmin1999} such that, except for the aforementioned viscous term, 
mechanical energy is conserved.  Interface deformation results then in a 
transfer, through the pressure jump, between surface and kinetic energy. In 
order to analyze such a transfer, we start by analyzing the evolution of kinetic 
energy for multiphase flows, which is obtained by taking the inner product of 
$\vel$ and, this time, the general formulation of an incompressible, Newtonian 
fluid given in equation (\ref{eqn:NS-multiphase})
\begin{equation}
  \frac{d E_k}{dt} = - \DOT{\vel}{\left( \rho \vel \cdot \nabla \right) \vel}
  - \DOT{\vel}{\grad p} + \DOT{\vel}{\div \deviatoric}
  \label{eqn:dEkdt-multi-continuum}
\end{equation}

As stated in Section~\ref{sec:DiffModel_single_conservation}, the skew-symmetry 
of the convective term results in a null contribution to kinetic energy, while 
the stress term includes an extra contribution due to the discontinuity at the 
interface stated in equation~(\ref{eqn:NS-multi-jumps}).
\begin{equation}
  - \DOT{\vel}{\grad p} + \DOT{\vel}{\div\deviatoric} =
    \DOT{\div \vel}{p}  - \DOT{\grad \vel}{\deviatoric}
      - \int_\Gamma \left[p\right]            \vel \cdot \ni dS
      + \int_\Gamma \left[\deviatoric\right]  \vel \cdot \ni dS
  \label{eqn:dEkdt-multi-0}
\end{equation}
Further details on the treatment of discontinuities within the inner product can 
be found in~\ref{sec:InnerProducts}.

Next, by considering an incompressible flow ($\div \vel = 0$), taking the 
pressure jump as stated in equation (\ref{eqn:NS-multiphase-jump}) and splitting 
$\nabla \vel$ into symmetric ($\strain$) and skew-symmetric ($\rotation$) parts 
we obtain
\begin{equation}
  \begin{split}
      \frac{d E_k}{dt} =
      - \DOT{\vel}{\grad p} + \DOT{\vel}{\div\deviatoric} &=
      - \DOT{\grad  \vel}{\deviatoric}
      - \int_\Gamma \vel \left[p\right]           \ni d\A
      + \int_\Gamma \vel \left[\deviatoric\right] \ni d\A\\
  &=- \DOT{\grad  \vel}{\deviatoric}
      + \st \int_\Gamma \ki \vel \cdot \ni d\A\\
  &=- \DOT{\strain + \rotation}{2\mu\strain}
      + \st \int_\Gamma \ki \vel \cdot \ni d\A\\
  &= -2\DOT{\strain}{\mu\strain}
      + \st \int_\Gamma \ki \vel \cdot \ni d\A\\
  &= -2\left\|\sqrt{\mu}\strain\right\|^2
      + \st \int_\Gamma \ki \vel \cdot \ni  d\A
  \qed
  \end{split}
  \label{eqn:dEkdt-multi-duality}
\end{equation}

As expected, viscosity results in a negative contribution to kinetic energy, 
whereas surface tension can take any sign depending on whether the interface is 
expanding or contracting.

On the other hand, the evolution of surface energy is related to the 
evolution of the interfacial area. By considering Helmholtz's free 
energy~\cite{Davies1961}
\begin{equation}
  dF = \st d\A,
  \label{eqn:HelmholtzFreeEnergy}
\end{equation}
and plugging it in equation (\ref{eqn:dSdt}), we state that
\begin{equation}
  \frac{dE_p}{dt}=  \int_\Gamma \frac{d}{dt}dF
                 =  \st \int_\Gamma \frac{d}{dt}d\A
                 = -\st \int_\Gamma \ki \vec{u} \cdot \ni d\A
  \label{eqn:dEpdt}
\end{equation}
Finally, performing a global balance of energy by combining equations 
(\ref{eqn:dEkdt-multi-duality}) and (\ref{eqn:dEpdt}), we obtain
\begin{equation}
    \frac{dE_m}{dt}
  = \frac{dE_k}{dt} + \frac{dE_p}{dt}
  = - 2\left\|\sqrt{\mu} S\right\|^2
  \label{eqn:dEtdt-final}
\end{equation}

As expected, surface tension does not play a role in the dissipation of energy, 
but rather produce a dynamic exchange between kinetic and surface ones of 
magnitude $\st\int_\Gamma \ki \vel \cdot \ni d\A$.

\subsection{Energy-preserving discretization}
\label{sec:DiffModel_multi_EP}
In the same spirit that symmetry-preserving methods aim at ensuring a null 
contribution of both pressure and convective terms in equation 
(\ref{eqn:dEkdt-single-continuum}) at the discrete level, the task in a 
multiphase flow simulation \replaced[id=A]{adds to the requirements}{is} to
preserve the proper transfer between kinetic and potential energies as
\begin{equation}
  \frac{d\sfield{E_m}}{dt} = \frac{d\sfield{E_k}}{dt} + \frac{d\sfield{E_p}}{dt}
  \label{eqn:TotalEnergyBalance}
\end{equation}
Namely, if symmetry-preserving schemes were constructed to satisfy at a discrete 
level equation (\ref{eqn:SymmPresGradientCondition}), energy-preserving methods 
\added[id=A]{also} satisfy \added[id=A]{ the discrete version of }equation 
(\ref{eqn:dSdt}) \deleted[id=A]{the discrete level} in order to properly capture 
energetic exchanges between kinetic and surface energies\added[id=R1]{.  This 
  requires the reformulation of the convective term
for variable density flows to preserve skew-symmetry, as proposed by Rozema et 
al.~\cite{Rozema2014}, which however is out of the scope of this work.  
Nonetheless, the transfer between kinetic and surface energy occurs trough the 
surface tension term}
as
\begin{equation}
  \frac{d}{dt}\DOT[F]{\GRAD \lsv{c}}{\ns} = - \DOT[F]{\U\GRAD \lsv{c}}{\ks}
  \label{eqn:GeomPressCondition}
\end{equation}
where $\U = \mathsf{diag}(\sfield[f]{u})\in \tsize{F}{F}$ is the diagonal 
arrangement of face velocities, $\lsv{c} \in \sfsize{C}$ is the cell-centered 
marker function vector and $\ks \in \sfsize{F}$ is the staggered curvature 
vector.  We consider the advection of the marker function in terms of the 
discretized equation (\ref{eqn:LS-advection})
\begin{equation}
  \frac{d\lsv{c}}{dt} = - \CONV[C]\lsv{c}
  \label{eqn:level-set-equation}
\end{equation}
where $\CONV[C] \in \tsize{C}{C}$ stands for the convective term of the marker 
function.  It may usually include a high-resolution scheme, as we shall see 
later, but so far we consider it as a single operator. We disregard the role of 
recompression stages in time derivatives but rather consider them as correction 
steps, which is discussed later on \added[id=R1]{this section}. As previously 
exposed for the continuum case, we can proceed by constructing the discrete 
counterpart of equation (\ref{eqn:dSdt-identity}) as
\begin{equation}
      \DOT[F]{\GRAD \frac{d \lsv{c}}{dt}}{\ns}
  = - \DOT[F]{\U\GRAD\lsv{c}}{\E \DIV \ns}
  \label{eqn:GeomPressDerivation}
\end{equation}
where a new shift operator, $\E \in \tsize{F}{C}$, is introduced in order to map 
the curvature from cells to faces. Exploiting the duality of the discrete 
gradient and divergence operators, equation 
(\ref{eqn:SymmPresGradientCondition}), we obtain
\begin{equation}
    -\DOT[C]{\frac{d \lsv{c}}{dt}}{\DIV\ns}
  = -\DOT[F]{\U\GRAD\lsv{c}}{\E \DIV \ns}
  \label{eqn:GeomPressDerivation0}
\end{equation}
By subsequently expanding the inner products, we obtain
\begin{equation}
    -\left( \frac{d \lsv{c}}{dt} \right)\T \metric{C} \DIV\ns
  = -\U\GRAD\lsv{c}\T \metric{F}  \E \DIV \ns \quad \forall \ns
  \label{eqn:GeomPressDerivation1}
\end{equation}
which must hold regardless of the interface normal, $\ns$, and consequently 
independently of the cell-centered curvature, $\DIV \ns$.  This implies that
\begin{equation}
  - \left( \frac{d \lsv{c}}{dt} \right)\T \metric{C}=
  - \left( \U \GRAD \lsv{c} \right)\T \metric{F} \E
  \label{eqn:GeomPressDerivation2}
\end{equation}
must hold at any time, while releasing a degree of freedom regarding the 
definition of the normal.
We can now plug equation (\ref{eqn:level-set-equation}) in for the time 
derivative and expand the transpose terms
\begin{equation}
  - \left( \frac{d \lsv{c}}{dt} \right)\T \metric{C}
  \overset{\deleted[id=R2]{(\ref{eqn:level-set-equation})}}{=}
    \left( \CONV[C] \lsv{c} \right)\T \metric{C}=
    \lsv{c}\T \CONV[C]\T \metric{C}
    \overset{\deleted[id=R2]{(\ref{eqn:GeomPressDerivation2})}}{=}
    - \lsv{c}\T \GRAD\T \U\T \metric{F} \E \quad \forall \lsv{c}
  \label{eqn:GeomPressDerivation3}
\end{equation}
which should hold for any $\lsv{c}$. This leads to
\begin{equation}
      \CONV[C]\T \metric{C}=
    - \GRAD\T \U\T \metric{F} \E
  \label{eqn:GeomPressDerivation4}
\end{equation}
where, exploiting the diagonal arrangement of both $\U$ and $\metric{F}$ to cast 
$\GRAD\T \U\T \metric{F}$ into $\GRAD\T \metric{F} \U$, we can use equation 
(\ref{eqn:DiscreteGradient}) to obtain the final condition as
\begin{equation}
  - \left( \metric{C} \CONV[C] \right)^T = \metric{C} \DIV \U \E
  \label{eqn:GeomPressConditionFinal}
\end{equation}

From where we can infer that the convective scheme of the marker function 
determines the curvature shift operator. This identity guarantees that energy 
transfers are balanced and thus total mechanical energy, $\sfield{E_m}$, is 
preserved up to temporal integration, in the same way that kinetic energy, 
$\sfield{E_k}$, is preserved in the symmetry-preserving discretization presented 
in Section~\ref{sec:DiffModel_single_SP} for the single-phase case. 

\begin{figure}[h]
  \centering
  \includegraphics[width=\textwidth]{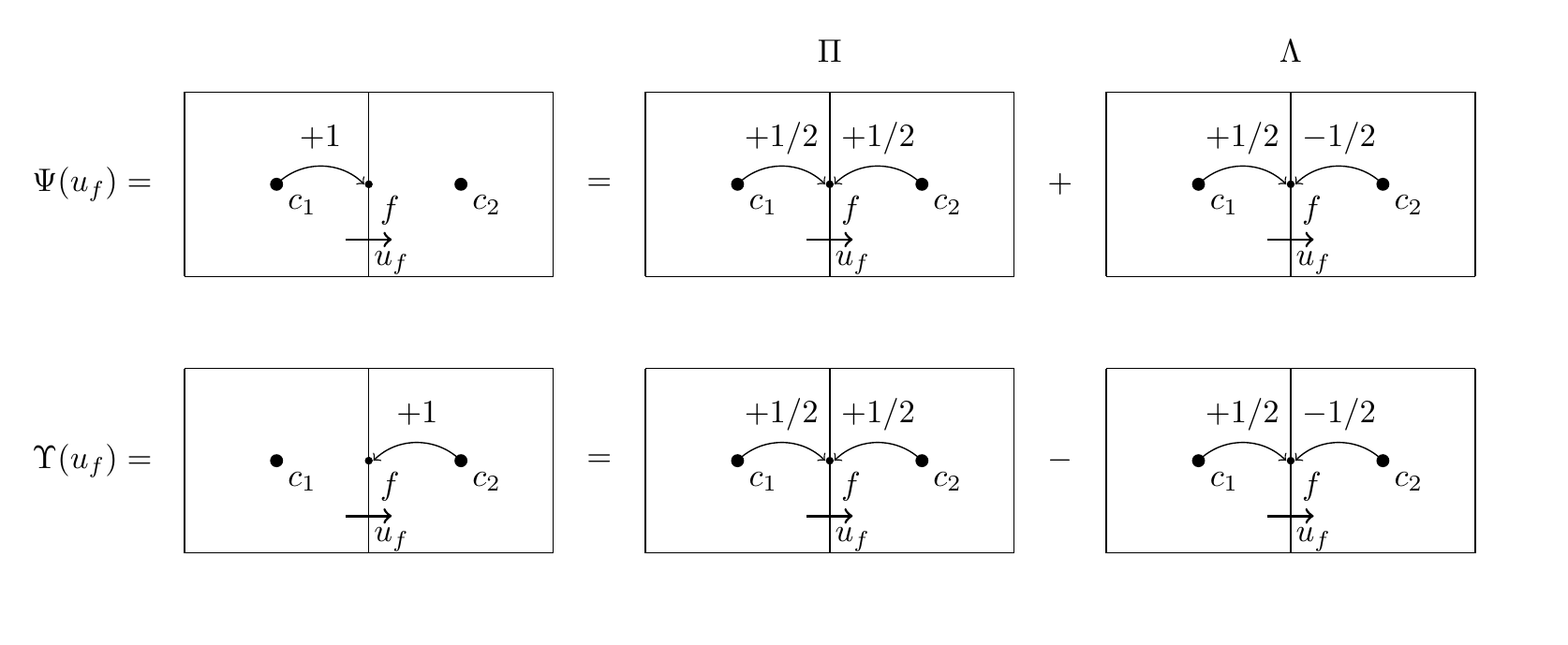}
  \caption{Example of a particular 
    high-\replaced[id=R2]{resolution}{ressolution} scheme $\HR$ for the 
  advection of $\lsv{c}$ (in this example, the well-known upwind scheme) and the 
corresponding dedicated curvature interpolator, $\E$.  In this case, the 
interpolation scheme for curvature is downwind.}
  \label{fig:EP-interpolator}
\end{figure}

Regarding the construction of $\CONV[C]$, any high-resolution scheme can be 
embedded into the algebraic form $\CONV[C]=\DIV\U\HR$, where 
$\HR\in\tsize{F}{C}$ is the actual high-resolution cell-to-face interpolator.  
For the CLS, this typically corresponds with SUPERBEE \cite{Olsson2005}.
We can split $\HR$ as $\HR=\sCF+\aCF$ \cite{Valle2018b}, to produce
\begin{equation}
  \CONV[C] = \DIV \U \left( \sCF + \aCF \right)
  \label{eqn:ConvectionSplit}
\end{equation}
This represents the symmetric ($\DIV \U \sCF$) and skew-symmetric ($\DIV \U 
\aCF$) components of $\CONV[C]$.
\added[id=R1]{ The extension to VOF schemes, nicely summarized by Patel et 
  al.~\cite{Patel2015a}, requires a previous casting of the advection scheme 
into the same framework introduced in~\cite{Valle2018b}.}
Plugging equation (\ref{eqn:ConvectionSplit}) into equation 
(\ref{eqn:GeomPressConditionFinal}) results in the final form of the dedicated 
cell-to-face interpolation for curvature
\begin{equation}
  \E = \sCF - \aCF
  \label{eqn:CurvatureInterpolator}
\end{equation}
which guarantees a proper potential and kinetic energy transfer. An
illustrative example can be seen in Figure \ref{fig:EP-interpolator}. In short, 
any upwind-like component used for the advection of $\lsv{c}$ turns into a 
downwind-like component for the interpolation of $\ks$. This can be compared 
with the second-order midpoint rule used by Olsson and Kreiss where 
$\E=\sCF$~\cite{Olsson2005}.

\subsection{Analysis}
\label{sec:Analysis-EP}
By mimicking equations (\ref{eqn:dEkdt-multi-continuum}) and 
(\ref{eqn:dEkdt-multi-duality}) we obtain the discrete counterpart of kinetic 
energy as
\begin{equation}
  \frac{d\sfield{E_k}}{dt}  = \st \DOT{\U\GRAD\lsv{c}}{\E \kc}
                            + \mu \DOT{\uf}{\LAP[F]\uf},
  \label{eqn:dEkdt-multi-discrete}
\end{equation}
which assumes a proper discretization of all other terms described in 
Section~\ref{sec:DiffModel_single}. We proceed similarly for potential energy by 
mimicking equation (\ref{eqn:dEpdt}) to define discrete potential energy as
\begin{equation}
  \frac{d\sfield{E_p}}{dt} = \st \DOT{\GRAD \frac{d\lsv{c}}{dt}}{\ns}
  \label{eqn:dEpdt-multi-discrete}
\end{equation}

We obtain the semi-discretized total energy equation by combining equations 
(\ref{eqn:dEkdt-multi-discrete}) and (\ref{eqn:dEpdt-multi-discrete}), which, in
combination with equation (\ref{eqn:GeomPressDerivation}) 
\replaced[id=R1]{yields}{yield}
\begin{equation}
  \frac{d\sfield{E_m}}{dt} =
                     \frac{d\sfield{E_k}}{dt} + \frac{d\sfield{E_p}}{dt}
                   = \st \DOT{\U\GRAD\lsv{c}}{\E \kc}
                   + \mu \DOT{\uf}{\LAP[F]\uf}
                   + \st \DOT{\GRAD \frac{d\lsv{c}}{dt}}{\ns}
                   = \mu \DOT{\uf}{\LAP[F]\uf}
                   \leq 0
  \label{eqn:dEmdt-multi-discrete}
\end{equation}

Which can be compared with equation (\ref{eqn:dEtdt-final}) to check that (in 
the absence of viscosity) the proposed numerical setup satisfies energy 
conservation up to temporal integration. Note that, as in the single-phase flow, 
this holds for any incompressible flow at the discrete level as well (i.e., 
$\DIV\uf=\Zv{c}$).
   
\added[id=R3, remark={Imported and reformulated from  
  Section~\ref{sec:Conclusions}}]{
The role of interface recompression deserves a special remark. Customarily 
included in the level set literature \cite{Sussman1994, Olsson2005}, its role is
to recover the interface sharpness that may have been deteriorated by the
convective schemes by taking additional, correcting steps after an initial 
advection stage. Nonetheless, even when performed conserving mass, as in 
\cite{Olsson2005, Olsson2007}, the nature of recompression results in a non-null
contribution to potential energy, which violates the conservation of mechanical 
energy. For this reason, the energy-preserving method presented here disregards 
recompression to focus on the physical coupling between marker advection and 
momentum transport.} \added[id=R1]{Similarly, other interface capturing schemes 
  may consider additional steps aimed at recovering interface quality and/or 
  mass conservation \cite{Ubbink1999}. While the results presented here allow to 
adopt this formulation into the momentum equation, including additional 
correcting steps require an individualized analysis.}

\section{Results}
\label{sec:Results}

Equipped with the discretization described in Section \ref{sec:DiffModel_multi}, 
we assess its performance for canonical tests for multiphase flow systems.  We 
focus on inviscid simulations in order to isolate the performance of our newly 
developed discretization. Equations (\ref{eqn:NS}) and (\ref{eqn:LS-advection}) 
are discretized according to the above-mentioned discretization. These read as
\begin{align}
  \label{eqn:discrete-momentum}
  \frac{d\uf}{dt} =& - \CONV[F]\uf - \GRAD \pc + \st \K[F] \GRAD \lsv{c} \\
  \label{eqn:discrete-marker}
  \frac{d\lsv{c}}{dt} =& - \CONV[C]\lsv{c}
\end{align}
where $\K[F] = \mathsf{diag}\left( \E\kc \right)$ is the diagonal arrangement of 
the staggered curvature.

Density ratio has been fixed to $1$ in order to isolate the surface tension 
term, simplify the discretization of the convective term and facilitate the 
solution of the pressure-velocity decoupling. Nonetheless, 
\deleted[id=R1]{although} as far as the convective term 
\replaced[id=R1]{preserves}{preserve} skew-symmetry and the Poisson equation is 
solved exactly, ratios different than $1$ may be included flawlessly.  Surface 
tension forces are included as mentioned in Section \ref{sec:DiffModel_multi}

The system is integrated in time with a second-order Adams-Bashforth scheme 
while the pressure-velocity decoupling is achieved with a classical FSM 
\cite{Chorin1968}. An efficient FFT decomposition in the periodic direction 
coupled with a Cholesky solver is used to ensure divergence-free velocity fields 
\deleted[id=A]{up} to machine accuracy.

All simulations are carried on a $\Omega = [2H\times2H]$ square domain, where 
$H$ is both the semi-width and semi-height of the cavity. Top and bottom faces 
present periodic boundary conditions, while at the sides no-flux boundary 
conditions is imposed for the marker function (i.e., $\grad \ls \cdot \nw = 0$) 
while free slip is set for velocity (i.e., $\vec{u} \cdot \nw =0$).  This 
enforces conservation of all physical quantities.

Linear perturbation theory is used to obtain reference values for time, velocity 
and pressure. Note that linear perturbation \replaced[id=R1]{assumes}{assume}
  small interfacial deformation, while the cases presented here do not 
  necessarily satisfy such a condition, it still provides with a reference 
  value.
Energy levels are scaled by and referenced to the initial observed mechanical 
energy. Because all simulations start with a fluid at rest and an elongated 
interface, kinetic energy evolves in the positive region (i.e., velocity is 
higher than or equal to the initial one) while potential energy evolves in the 
negative region (i.e., elongation is less than or equal to the initial one).

Tests are carried in order to compare the standard midpoint rule used for the 
interpolation of curvature proposed by Olsson and Kreiss \cite{Olsson2005} with 
the newly developed interpolation scheme.
Recompression has been initially set aside in order to evaluate its impact on 
both schemes in a subsequent analysis. It is computed as
\begin{equation}
  \frac{d\lsv{c}}{d\tau} + \DIV \Gamma_{f \to c} \Nc \left( \Ic{C} - \LS \right) 
  \lsv{c} =  \DIV \sfield[F]{E} \GRAD \lsv{c},
  \label{eqn:CLS-recompression}
\end{equation}
where $\tau$ stands for pseudo-time, $\Gamma_{f \to c}$ is vector-valued shift 
operator, $\Nc \in \tsize{dC}{C}$ maps scalars to vector fields aligned with the 
interface normal, while
$\LS = \mathsf{diag}(\lsv{c})$ and $\sfield[F]{E} = 
\mathsf{diag}(\sfield[f]{\upepsilon})$ are the diagonal arrangements of, 
respectively, $\lsv{c}$ and $\epsilon$; where $\epsilon$ is the face-centered 
smoothing factor defined in Section~\ref{sec:Model_Topology_Capturing}.
Further details can be found in Olsson and Kreiss~\cite{Olsson2005} for the CLS 
and in Trias et al.~\cite{Trias2014} for the construction of the operators.

\ifStaticBubble
\subsection{Cylindrical column}
The classical setup of a zero gravity cylindrical column of liquid is tested in 
order to show the impact of the newly proposed method into spurious currents.  
The section of the column is located at the center of the domain and is given a 
radius of $R_0 = 0.3H$. Velocity is initially stagnant and that is how it should 
remain throughout the simulation; however, spurious currents are expected to 
appear due to errors in the calculation of curvature \cite{Magnini2016}. The 
initial setup is depicted in Figure~\ref{fig:StaticBubble-InitialSetup}.

Linear perturbation theory provides with the time period for an initially 
cylindrical interface perturbed as $r(\theta)=R_0+r_pcos(s\theta)$, where
$s=2,3,4,\dots$ \replaced[id=R1]{corresponding to}{correspond with} ellipsoidal,
triangular of rectangular deformations, respectively \cite{Lamb1945}. Because 
linear theory predicts perfect equilibrium for both $s=0$ and $s=1$, we 
arbitrarily assume an ellipsoidal perturbation (i.e., $s=2$) in order to obtain 
a reference state. The oscillation period can be computed for any $s$ as
$T = 2\pi/\sqrt{2\rho R_0^3/\gamma s (s^2-1)}$ \cite{Fyfe1988}, while the 
characteristic length scale is $L=2\pi R_0$, which leads a characteristic speed 
of $c = L/T = \sqrt{\gamma s (s^2-1)/2\rho R_0^3}$, while pressure is referenced 
to $\rho c^2$. Integration is carried over $5T$.

\begin{figure}[h!]
  \centering
  \includegraphics[width=5cm, trim={4cm 1cm 0cm 4cm},clip]
  {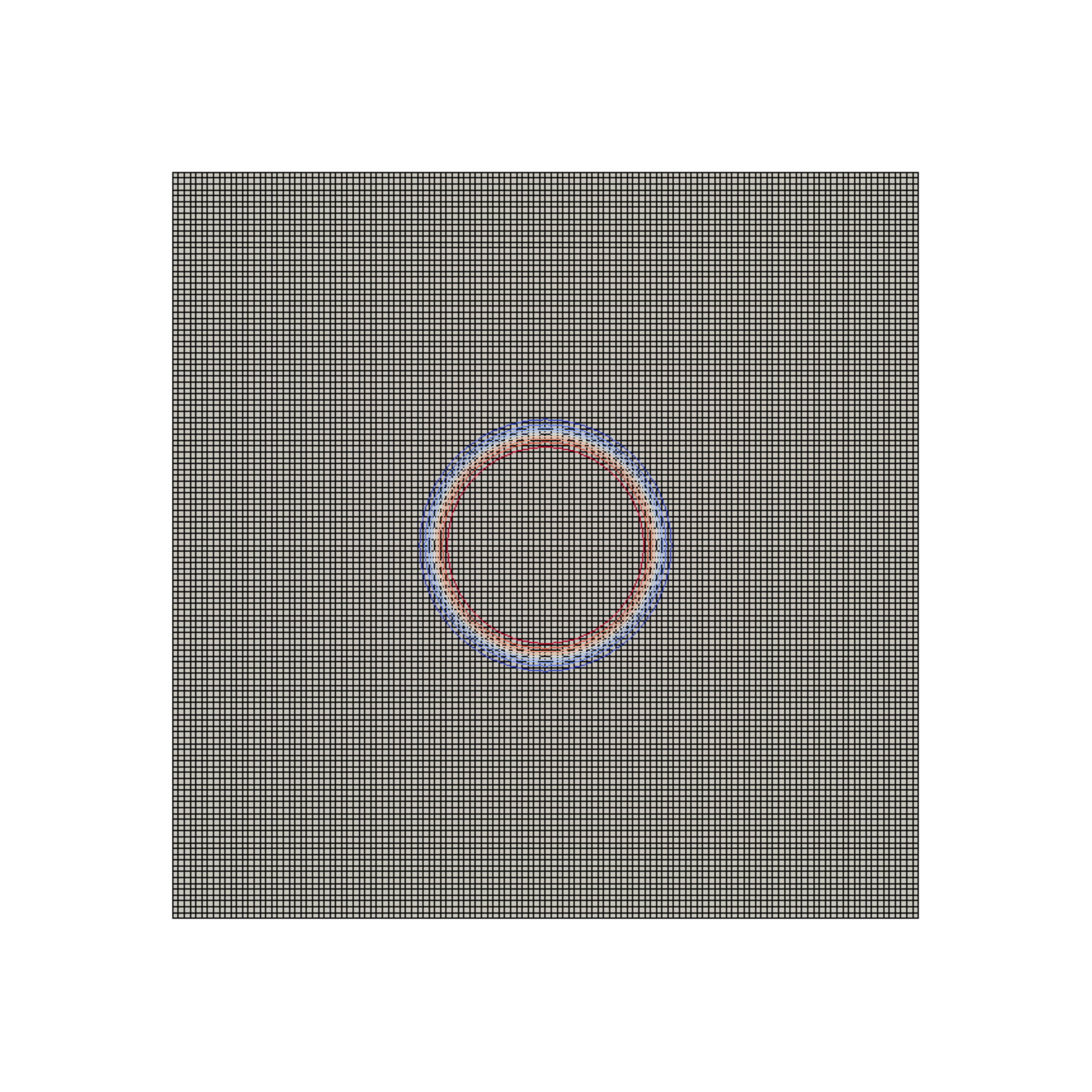}
  \caption{Initial setup of the marker function for cylindrical column test case 
  in a $128\times128$ mesh. Contour lines are plotted for $\Delta\ls = 0.1$.}
  \label{fig:StaticBubble-InitialSetup}
\end{figure}

Results in Figure~\ref{fig:EnergyEvolution-bubble-adv} show how the newly 
proposed method (right column) results in \replaced[id=R1]{an}{a} energy stable 
simulation by counterbalancing the numerical increase in kinetic energy with a 
decrease of potential energy. This yields to a stagnant situation in which both 
kinetic and potential energy restore their initial values (i.e., $E_k=0$ and 
$E_p=0$).  On the other hand, the standard midpoint rule interpolation for 
curvature (left column) results in an increase in total energy \deleted[id=R1]{, 
which compromises the stability of the system in the long run}.

In Figure~\ref{fig:bubble-fields-0-128-41} it can be seen how the newly 
developed curvature interpolation scheme (right) provides, first of all, an 
order of magnitude smaller oscillations that the standard one produces (left).  
In addition, there is a dramatic increase in the flow quality within the 
interface, extending the benefits of the high resolution advection scheme for 
the marker into the velocity field. On the other hand, the use of the standard 
midpoint rule for updating curvature pollutes the flow within both phases.

\begin{figure}[htpb]
  \centering
  \includegraphics[width=\textwidth]{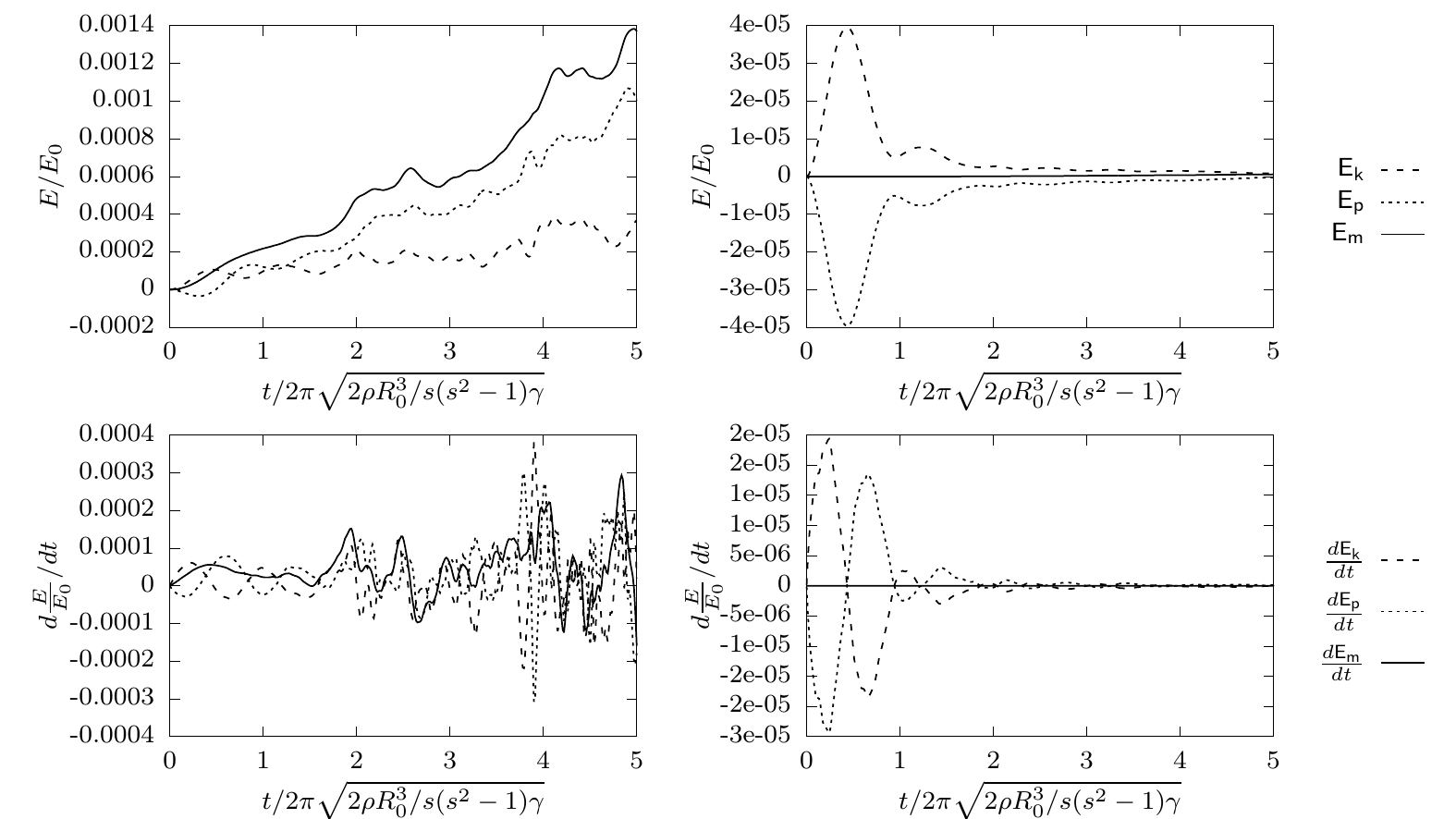}
  \caption{Energy evolution of the cylindrical section for a pure advection case 
    (i.e., no recompression) with the standard interpolation of Olsson and 
    Kreiss \cite{Olsson2005} for the curvature (left) and the newly proposed 
    method (right) in a $128\times128$ mesh. Top rows show the discrete values 
    of kinetic ($\sfield{E_k}$), potential ($\sfield{E_p}$) and total 
    ($\sfield{E_m}$) energy.  Bottom rows show their semi-discretized time 
    derivative according to equations (\ref{eqn:dEkdt-multi-discrete}), 
  (\ref{eqn:dEpdt-multi-discrete}) and (\ref{eqn:dEmdt-multi-discrete}), 
respectively.}
  \label{fig:EnergyEvolution-bubble-adv}
\end{figure}
\begin{figure}[htpb]
  \centering
  \includegraphics[width=0.49\textwidth, trim={4cm 1cm 0cm 4cm} ,clip] 
  {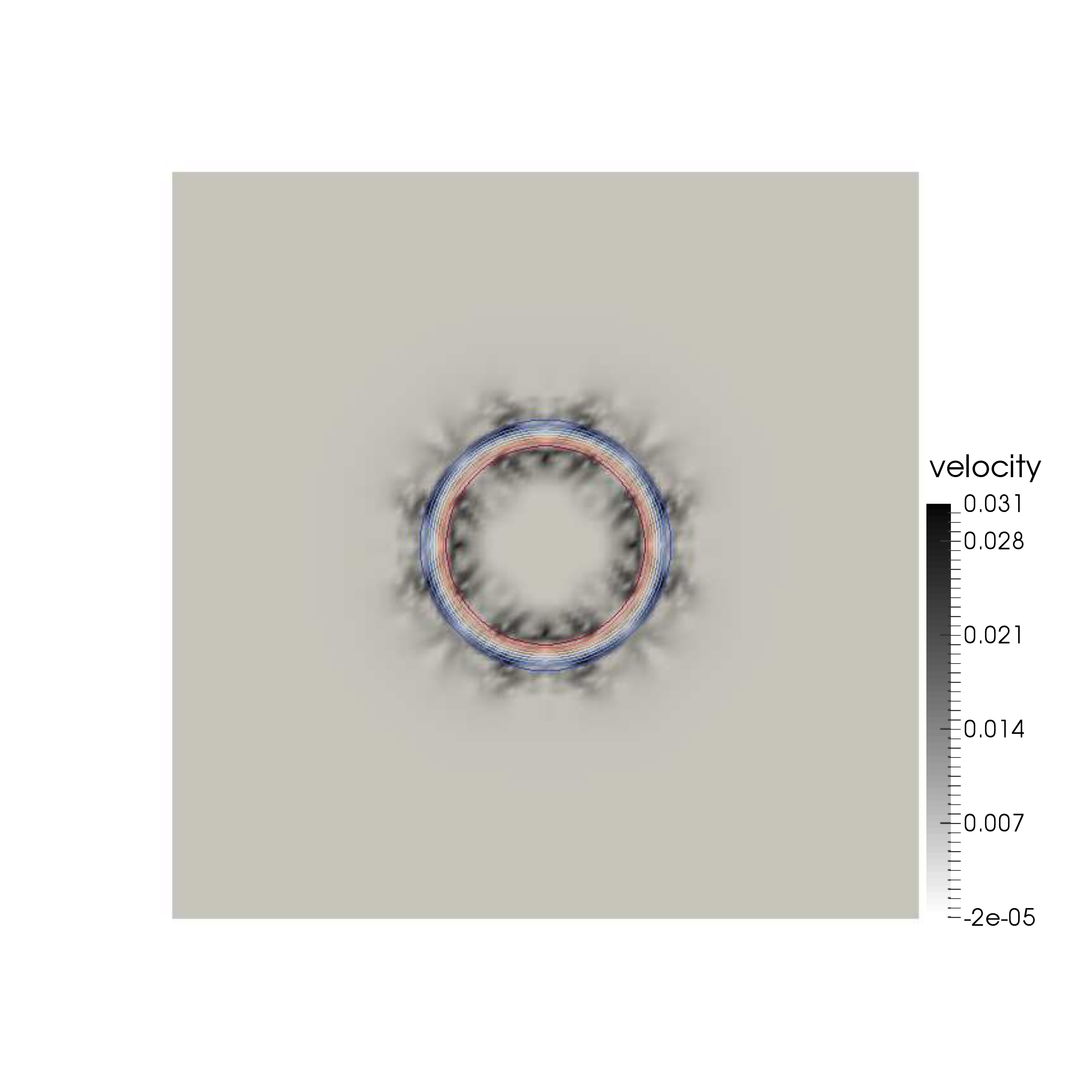}
  \includegraphics[width=0.49\textwidth, trim={4cm 1cm 0cm 4cm} ,clip] 
  {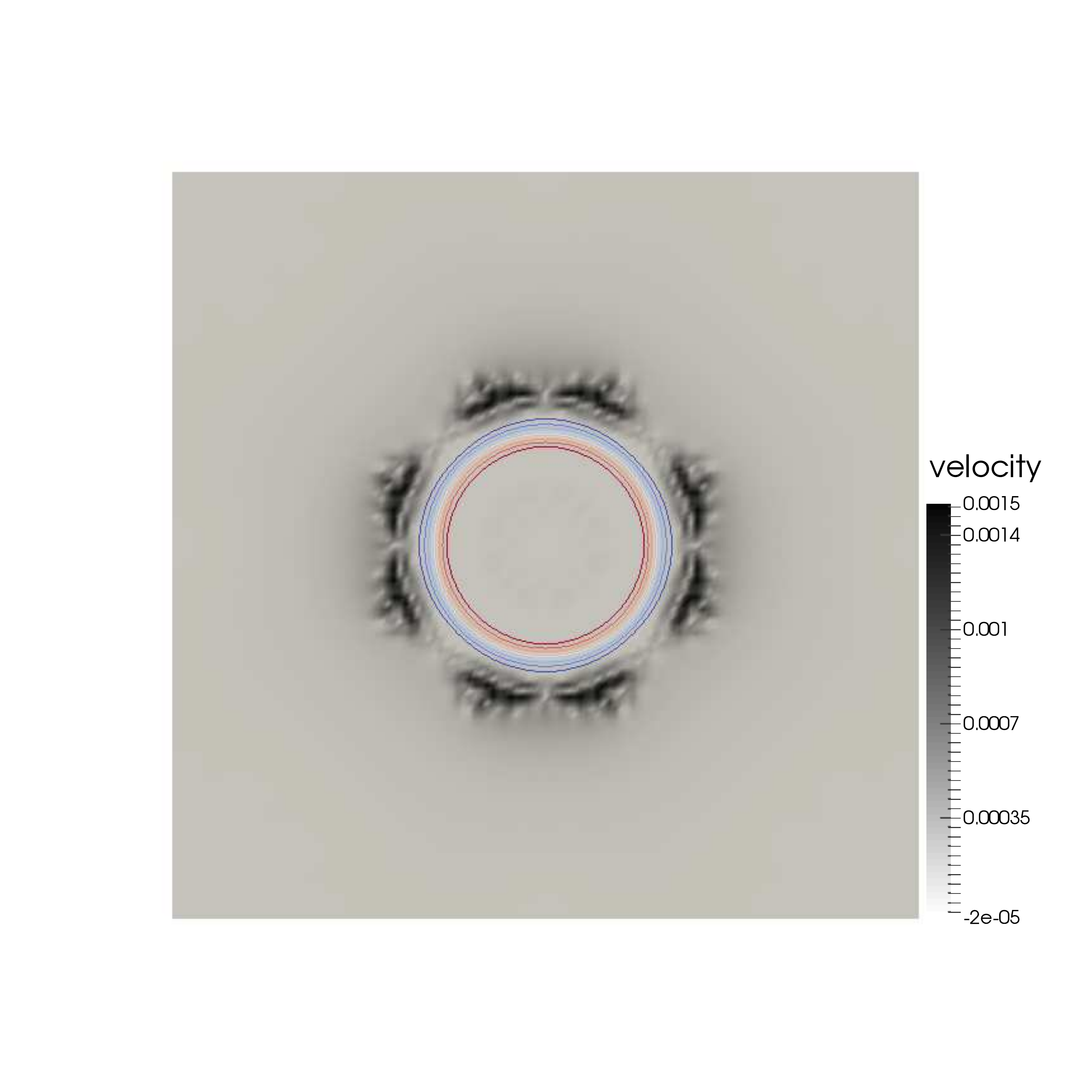}
  \caption{Velocity magnitude and interface location at $t = 5T$ for a 
   cylindrical column in a $128\times128$ mesh advected without recompression.  
   Left figure uses the standard midpoint rule while the right one uses the 
   newly developed energy-preserving one. Contour lines are plotted for 
   $\Delta\ls =  
 0.1$.}
  \label{fig:bubble-fields-0-128-41}
\end{figure}

It is remarkable how, despite initializing the interface to a theoretical 
minimum energy situation (i.e., cylindrical cross-section), numerical imbalances 
when computing curvature does not reflect such a situation \cite{Magnini2016}.  
Nonetheless, the use of an energy-preserving scheme acts in order to keep energy 
constant, and so counter-balances such an artificial movement by modifying the 
curvature accordingly. This results in a robust method which eventually is 
perturbation-proof.

\begin{figure}[h]
  \centering
  \includegraphics[width=\textwidth]{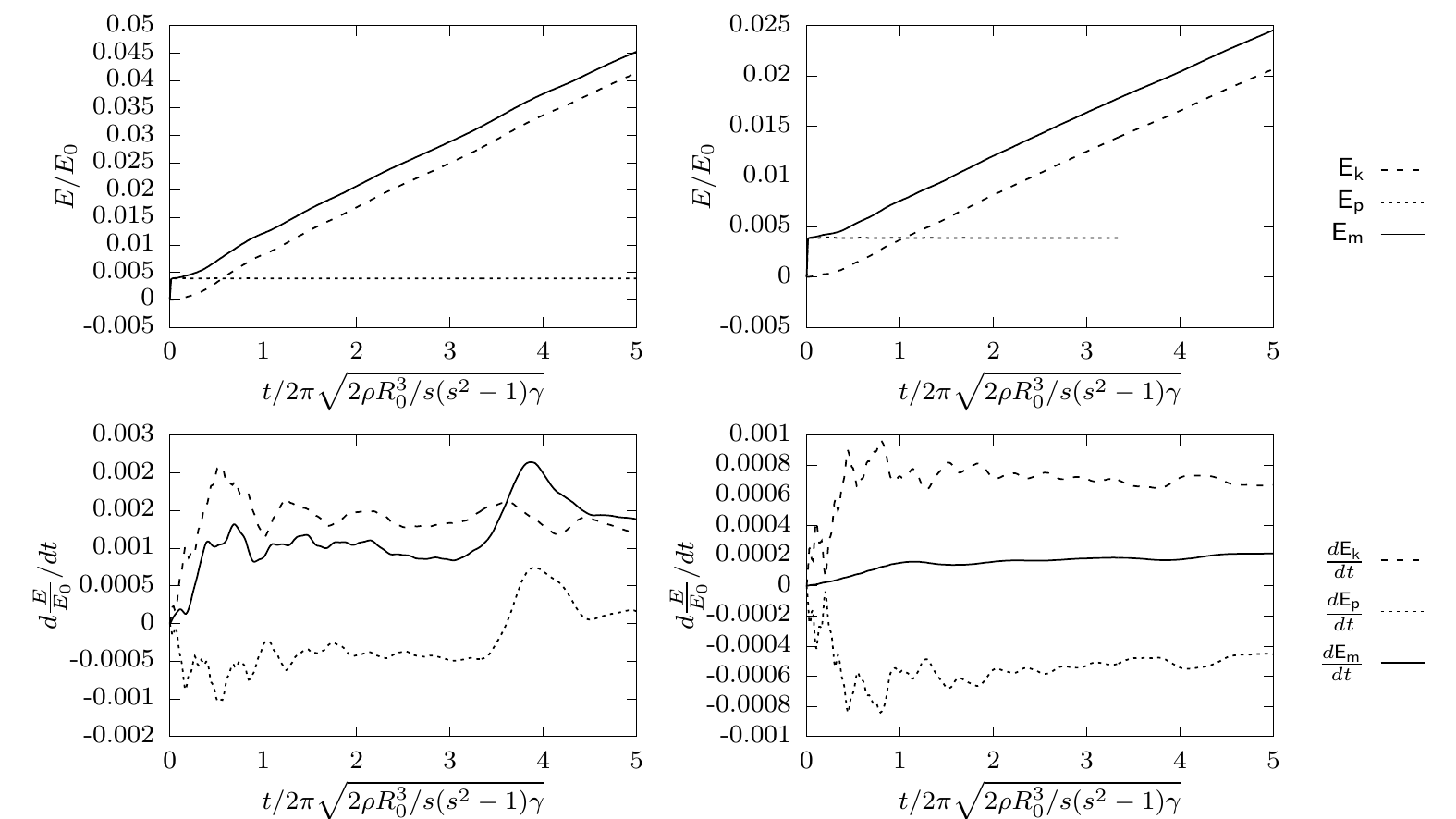}
  \caption{Energy evolution of cylindrical column with the complete Olsson and 
    Kreiss method \cite{Olsson2005} with a single recompression stage (left), 
    and the same method including the modified curvature interpolation (right) 
    in a $128\times128$ mesh. Top rows show the discrete values of kinetic 
    ($\sfield{E_k}$), potential ($\sfield{E_p}$) and total ($\sfield{E_m}$) 
    energy.  Bottom rows show their semi-discretized time derivative according 
    to equations (\ref{eqn:dEkdt-multi-discrete}), 
    (\ref{eqn:dEpdt-multi-discrete}) and (\ref{eqn:dEmdt-multi-discrete}), 
  respectively.}
  \label{fig:EnergyEvolution-bubble-rec}
\end{figure}
\begin{figure}[h]
  \centering
  \includegraphics[width=0.49\textwidth, trim={4cm 1cm 0cm 4cm} ,clip] 
  {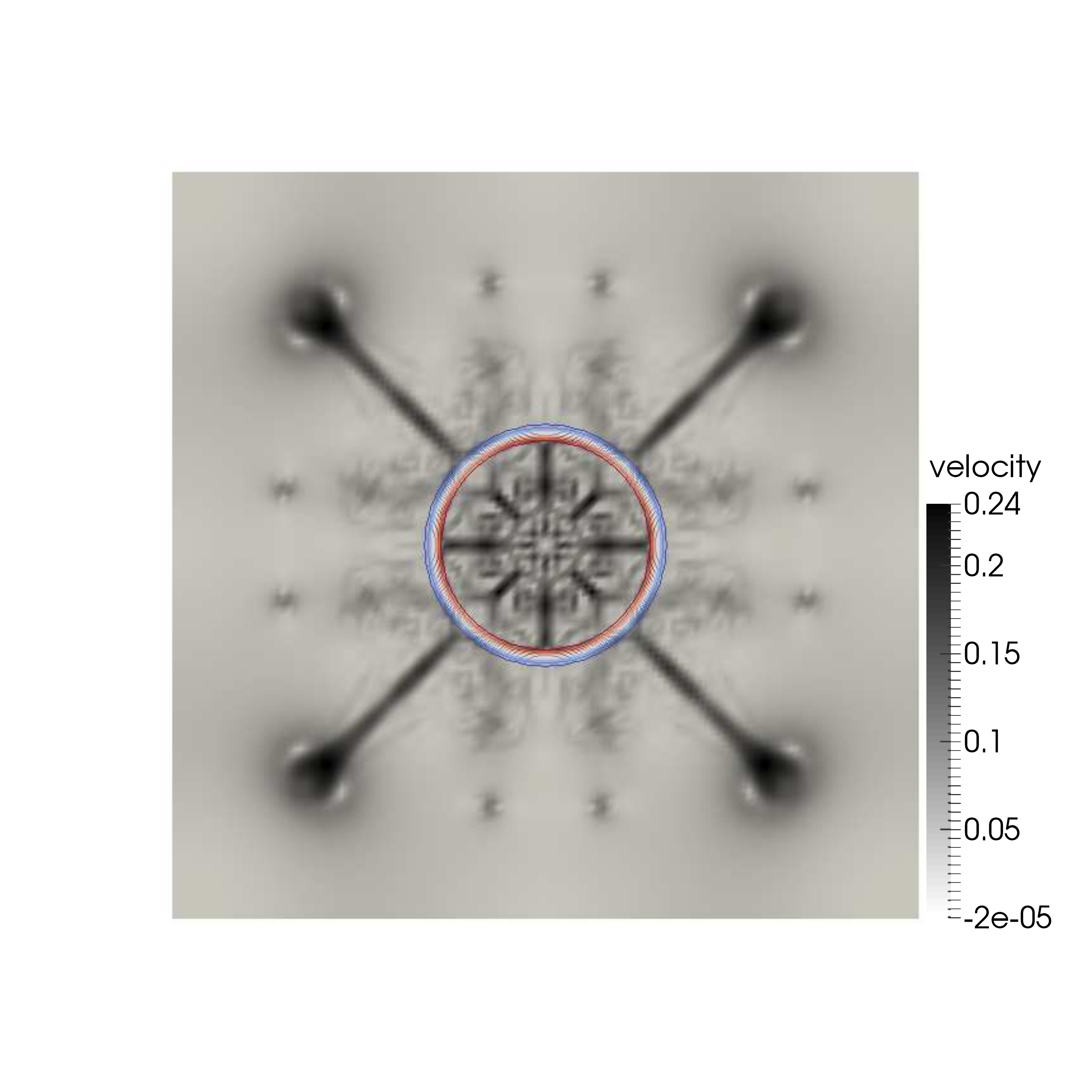}
  \includegraphics[width=0.49\textwidth, trim={4cm 1cm 0cm 4cm} ,clip] 
  {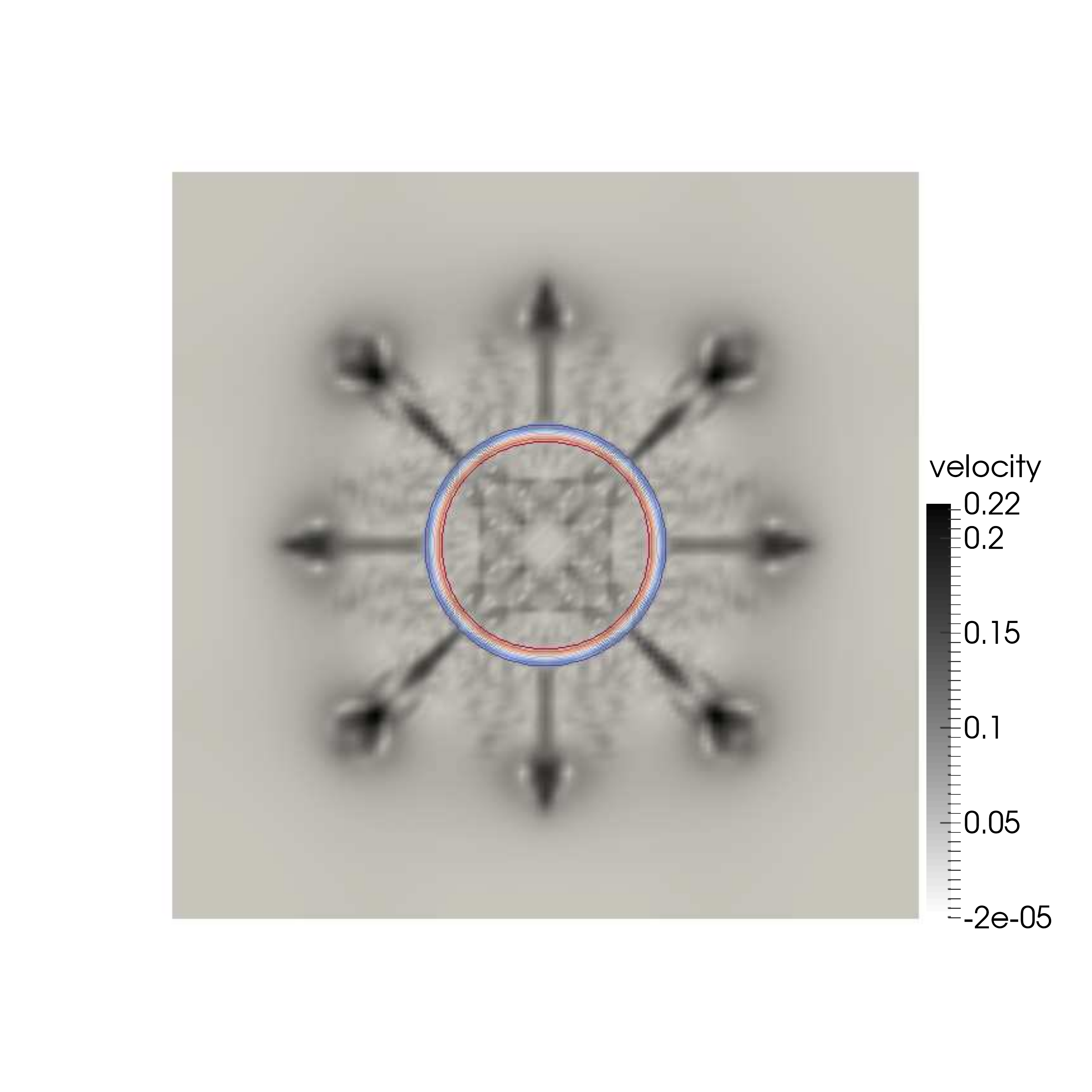}
  \caption{Velocity magnitude and interface location at $t = 5T$ for a 
   cylindrical column in a $128\times128$ mesh advected with a single 
   recompression step.  Left figure uses the standard 2nd order shift operator 
   while the right one uses the newly developed energy-preserving one.  Contour 
 lines are plotted for $\Delta\ls = 0.1$.}
  \label{fig:bubble-fields-1-128-41}
\end{figure}

In comparison with
Figure~\ref{fig:EnergyEvolution-bubble-adv},
Figure~\ref{fig:EnergyEvolution-bubble-rec} shows the impact of recompression
in both schemes. As can be seen, the newly developed interpolation method can do 
little in terms of energy, as the recompression stage
\replaced[id=A]{increases}{increase} the energy of the system. Actually, we see 
how the increase in kinetic energy is even higher than in the previous case, 
with no recompression associated.

On the other hand, Figure~\ref{fig:bubble-fields-1-128-41} shows how the 
velocity field is clearly distorted in both cases, degrading the solution with 
respect to the pure advection algorithm one and two orders of magnitude with 
respect to the midpoint and the energy-preserving interpolation schemes, 
respectively.  Noticeably, we still retain, even by including the recompression 
scheme, a higher quality of the velocity field within the bounded region for the 
newly developed interpolation scheme. The impact of recompression in the overall 
quality of the solution is discussed in Section \ref{sec:Conclusions}.
\FloatBarrier
\fi
\ifOscillatingEllipse
\subsection{Oscillating ellipsoidal column}
In order to stretch the previous result to a dynamic equilibrium situation, an 
ellipsoidal section is set by distorting the initially cylindrical case.
As in the cylindrical water column, spurious currents may appear, while this 
time they accompany legitimate currents as a result of
regions with a moderate non-constant curvature.
The ellipse is centered in the domain and is defined by $x = 0.5 cos(\alpha)$ 
and $y=0.3sin(\alpha)$, where $\alpha \in [0,2\pi)$. Velocity field is 
initialized at rest and should follow to the oscillation of the ellipsoid 
throughout the simulation. The initial setup is depicted in 
Figure~\ref{fig:OscillatingEllipse-InitialSetup}.

In the same fashion that in the cylindrical water column described above, linear 
perturbation theory is employed in order to obtain a reference state.  
Characteristic length is set to $L=2\pi R_0$, where $R_0 = 0.3$. Time,
velocity and pressure scales used are the same than those for the cylindrical 
section case.

\begin{figure}[h]
  \centering
  \includegraphics[width=5cm, trim={4cm 4cm 2cm 4cm},clip]
  {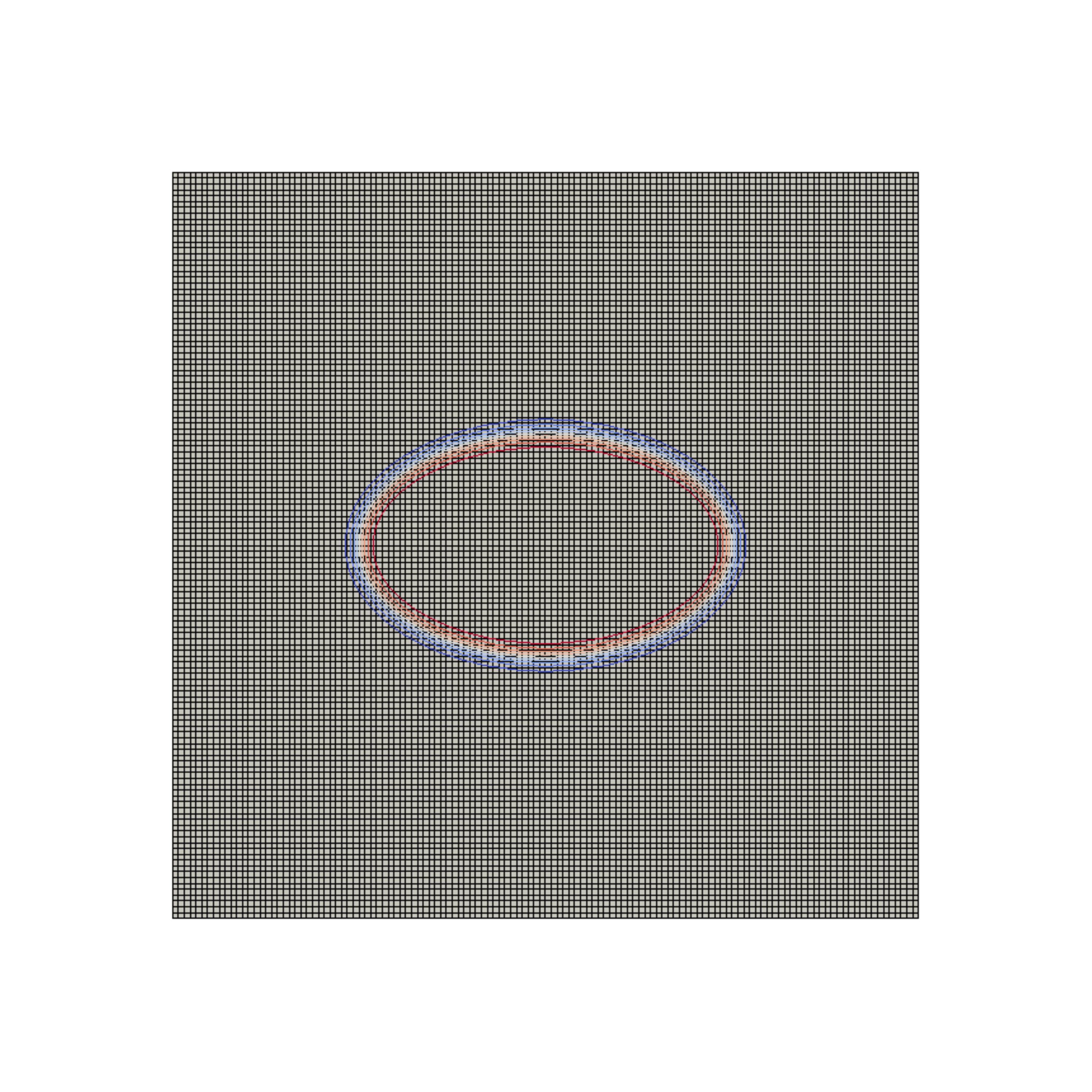}
  \caption{Initial setup of the marker function for the oscillating ellipse test 
  case in a $128\times128$ mesh. Contour lines are plotted for $\Delta\ls = 
0.1$.}
  \label{fig:OscillatingEllipse-InitialSetup}
\end{figure}

Figure~\ref{fig:EnergyEvolution-ellipse-adv} shows how, while the standard 
midpoint interpolation (left) clearly \replaced[id=R1]{increases}{increase} the 
mechanical energy of the system, the newly proposed energy-preserving 
interpolation scheme for curvature (right) preserves mechanical energy, which 
yields physically consistent results and numerically stable simulations. There 
is, however, both positive and negative offsets for kinetic and potential 
energies. While kinetic and potential energy are supposed to oscillate between 
$0$ and its maximum or minimum for an ideal harmonic oscillator, we observe that 
this is not the case.  This is explained by an imbalance in the momentum 
equation, which provides an artificial acceleration in the fluid, resulting in 
an increase of the kinetic energy base state \cite{Popinet2017}. By virtue of 
the energy-preserving scheme the oscillation gap for potential energy is reduced 
accordingly, resulting in a decrease of the elongation amplitude. This plays a 
relevant role in the next case presented, the capillary wave, which is further 
discussed in the next subsection. Despite this well-known issue results still 
show the expected oscillatory behavior of the ellipsoid. This can be checked 
from the bottom row of Figure~\ref{fig:EnergyEvolution-ellipse-adv}, where the 
magnitude of the energy transfers \replaced[id=R1]{remains}{remain} 
approximately constant \replaced[id=R1]{throughout}{along} the simulation. In 
terms of the oscillating behavior, the increase in mechanical energy for the 
naive interpolation results not only in artificially higher values of kinetic 
energy, but also in a phase difference with respect to the energy-preserving 
one.

Figure~\ref{fig:ellipse-fields-0-128-41} \replaced[id=R1]{presents}{present} the
marker and velocity fields after $t=5T$ with a pure advection scheme. Results 
for the energy-preserving scheme (right) show a shift in phase with respect to 
the midpoint interpolation scheme (left). Velocity is not only higher for the 
naive approach, but also the shape of the interface provides with non-physical 
curvature, as it can be observed by the kink appearing along the horizontal 
centerline of the ellipsoid (left), which can be compared with the smoother 
profile present in the energy-preserving approach (right).

\begin{figure}[h]
  \centering
  \includegraphics[width=\textwidth]{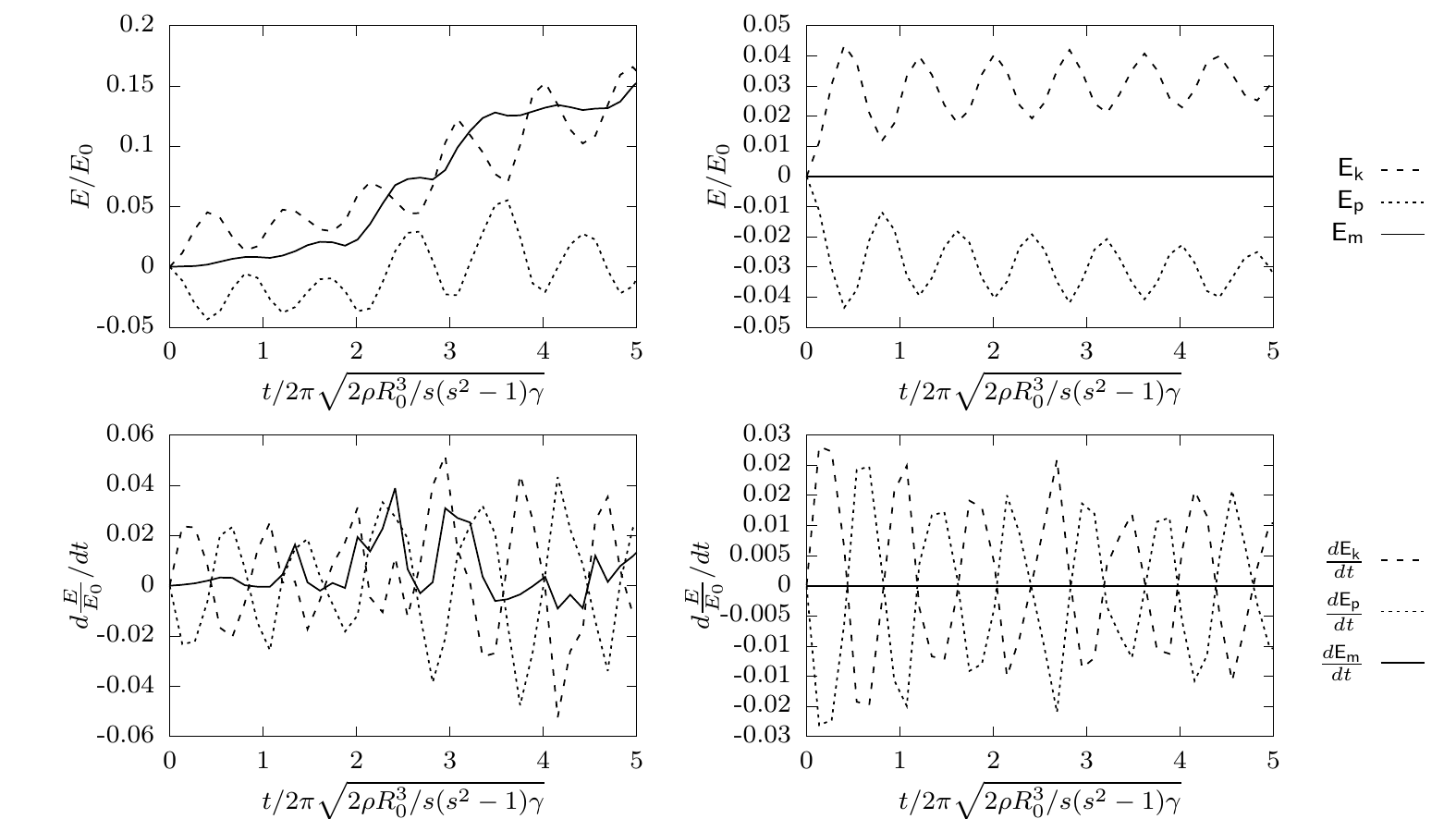}
  \caption{Energy evolution of the ellipsoidal section for a pure advection case 
    (i.e., no recompression) with the standard interpolation of Olsson and 
    Kreiss \cite{Olsson2005} for the curvature (left) and the newly proposed 
    method (right) in a $128\times128$ mesh. Top rows show the discrete values 
    of kinetic ($\sfield{E_k}$), potential ($\sfield{E_p}$) and total 
    ($\sfield{E_m}$) energy.  Bottom rows show their semi-discretized time 
    derivative according to equations (\ref{eqn:dEkdt-multi-discrete}), 
  (\ref{eqn:dEpdt-multi-discrete}) and (\ref{eqn:dEmdt-multi-discrete}), 
respectively.}
  \label{fig:EnergyEvolution-ellipse-adv}
\end{figure}
\begin{figure}[h]
  \centering
  \includegraphics[width=0.49\textwidth, trim={4cm 1cm 0cm 4cm} ,clip] 
  {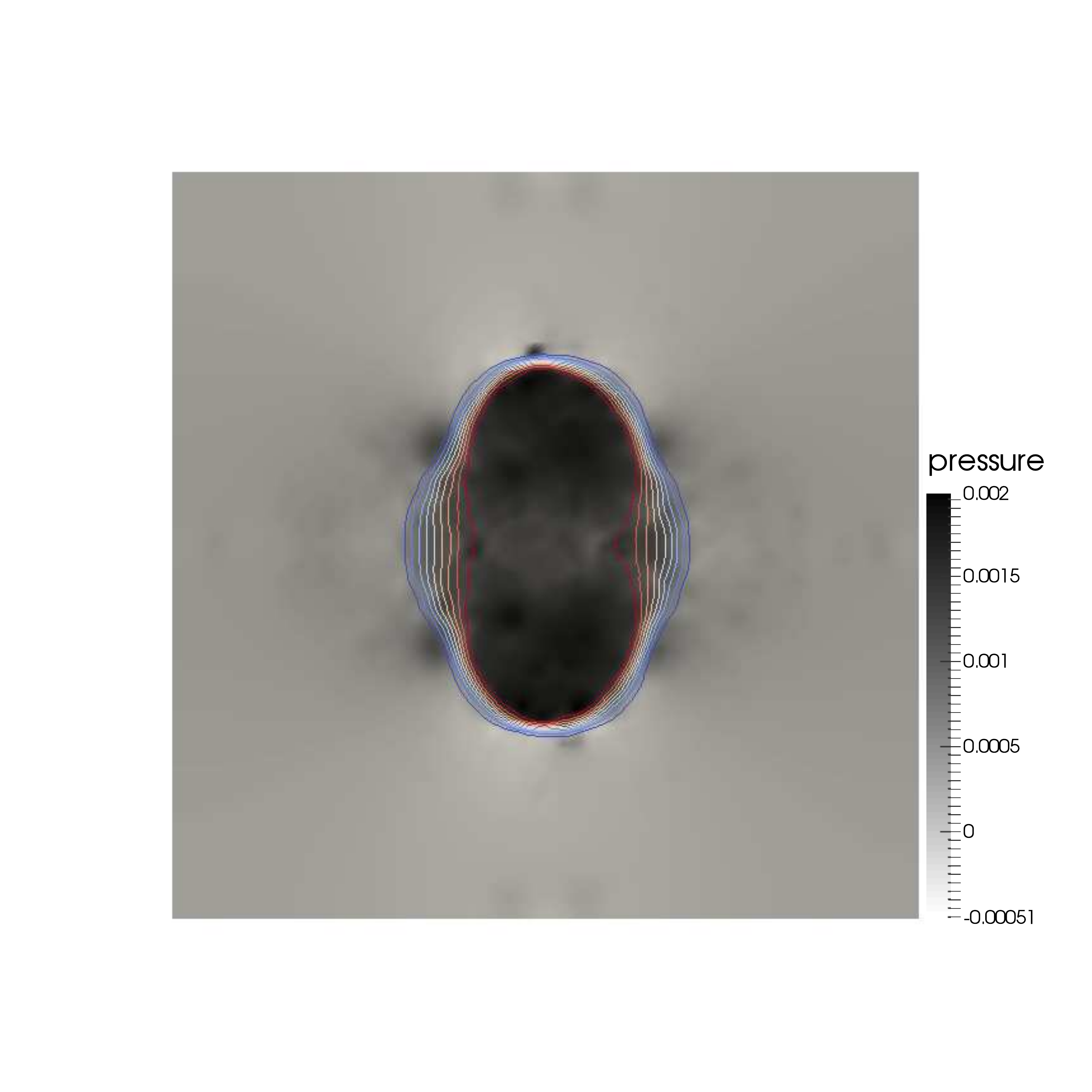}
  \includegraphics[width=0.49\textwidth, trim={4cm 1cm 0cm 4cm} ,clip] 
  {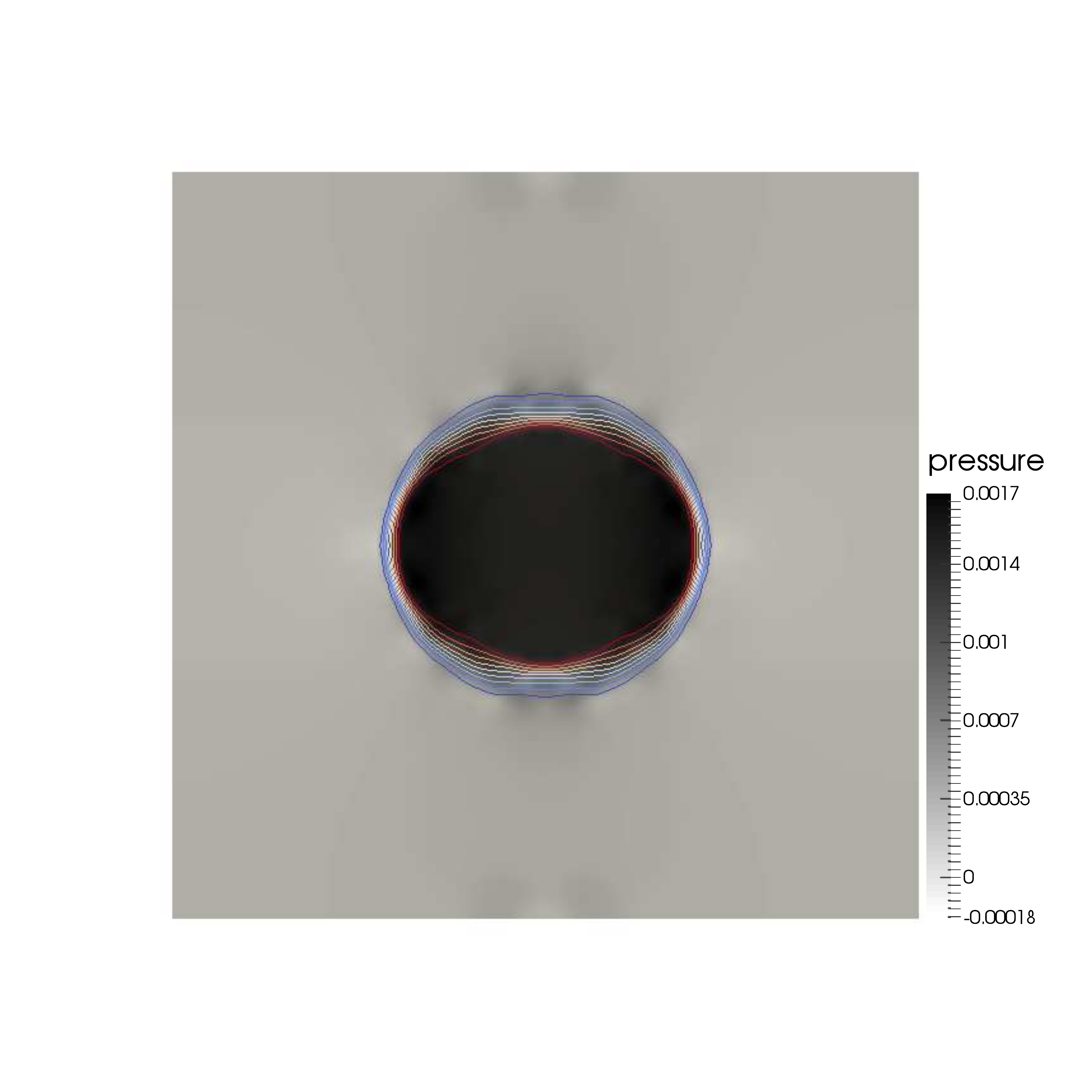}
  \caption{Pressure field and interface location at $t = 5T$ for the oscillating 
  ellipse in a $128\times128$ mesh advected without recompression. Left figure 
  uses the standard 2nd order shift operator while the right one uses the newly 
  developed energy-preserving one.  Contour lines are plotted for $\Delta\ls = 
  0.1$.}
  \label{fig:ellipse-fields-0-128-41}
\end{figure}

In summary, the use of the energy-preserving scheme 
\replaced[id=R1]{provides}{present with} a higher degree of reliability, by 
preserving mechanical energy also in a dynamical equilibrium situation.  Despite
the numerical errors in which the discretization of momentum may occur, the 
method is robust and still preserves mechanical energy.

\begin{figure}[h]
  \centering
  \includegraphics[width=\textwidth]{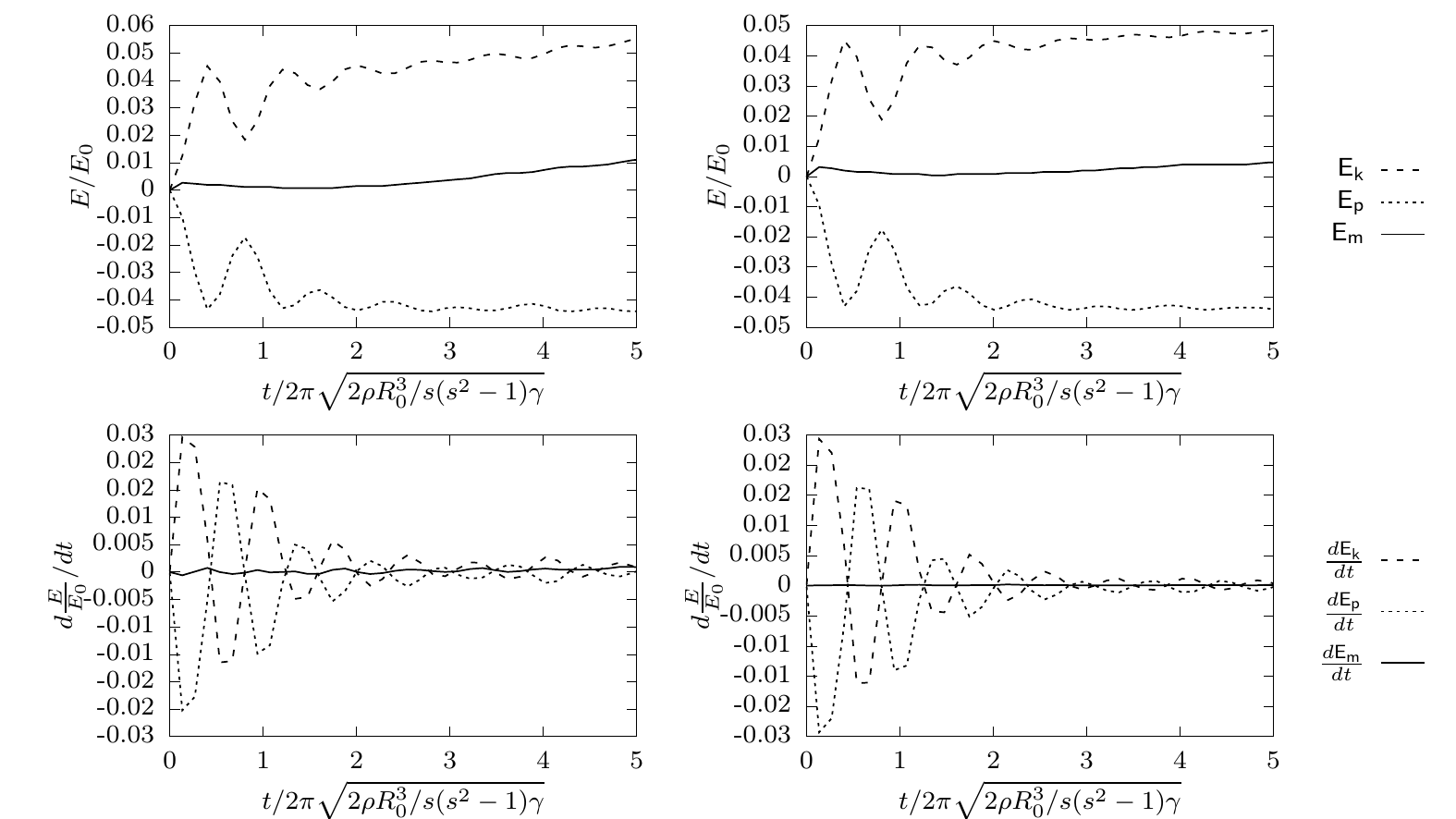}
  \caption{Energy evolution of the oscillating ellipse with the complete Olsson 
    and Kreiss method \cite{Olsson2005} with a single recompression stage 
    (left), and the same method including the modified curvature interpolation 
    (right) in a $128\times128$ mesh. Top rows show the discrete values of 
    kinetic ($\sfield{E_k}$), potential ($\sfield{E_p}$) and total 
    ($\sfield{E_m}$) energy.  Bottom rows show their semi-discretized time 
  derivative according to equations (\ref{eqn:dEkdt-multi-discrete}), 
(\ref{eqn:dEpdt-multi-discrete}) and (\ref{eqn:dEmdt-multi-discrete}), 
respectively.}
  \label{fig:EnergyEvolution-ellipse-rec}
\end{figure}
\begin{figure}[h]
  \centering
  \includegraphics[width=0.49\textwidth, trim={4cm 1cm 0cm 4cm} ,clip] 
  {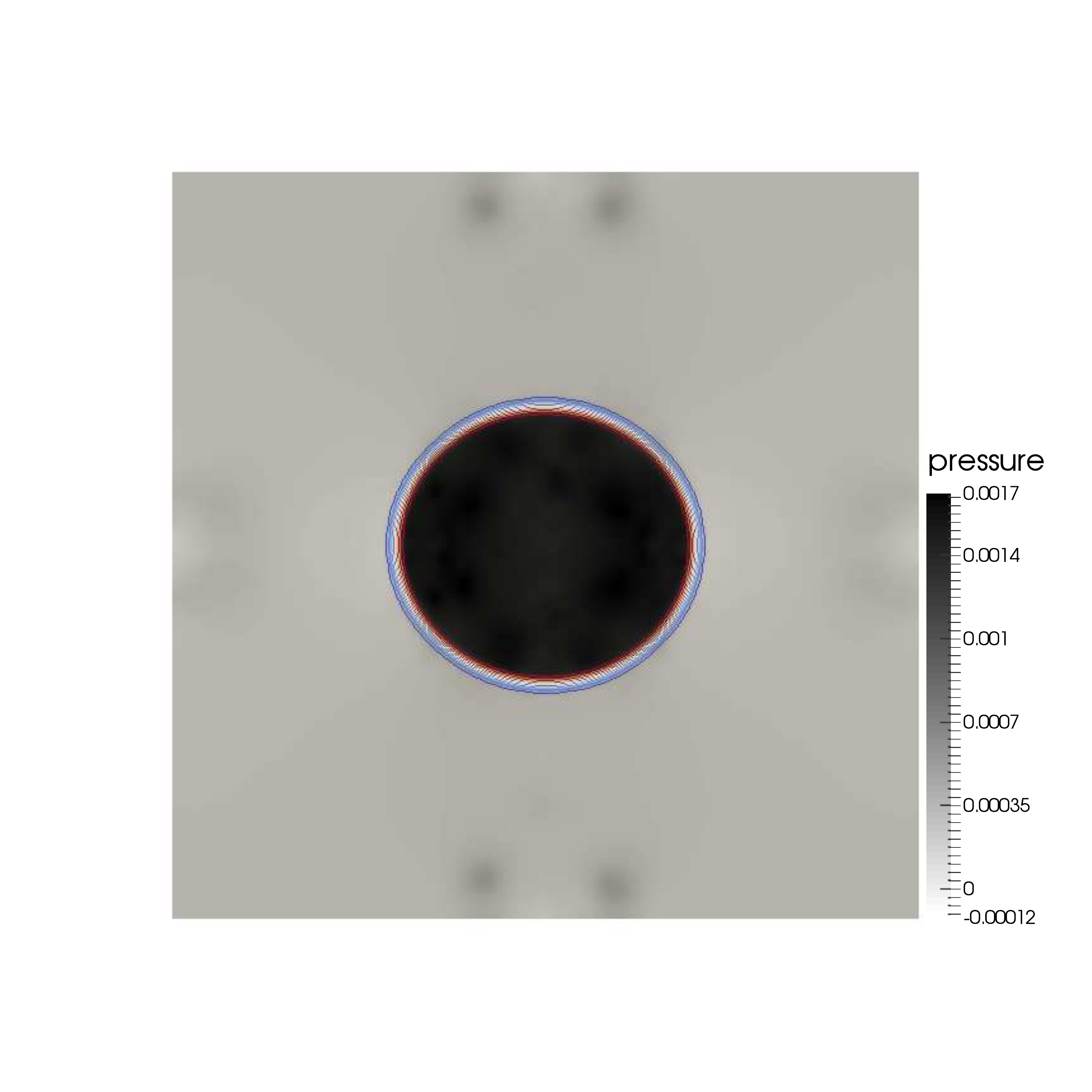}
  \includegraphics[width=0.49\textwidth, trim={4cm 1cm 0cm 4cm} ,clip] 
  {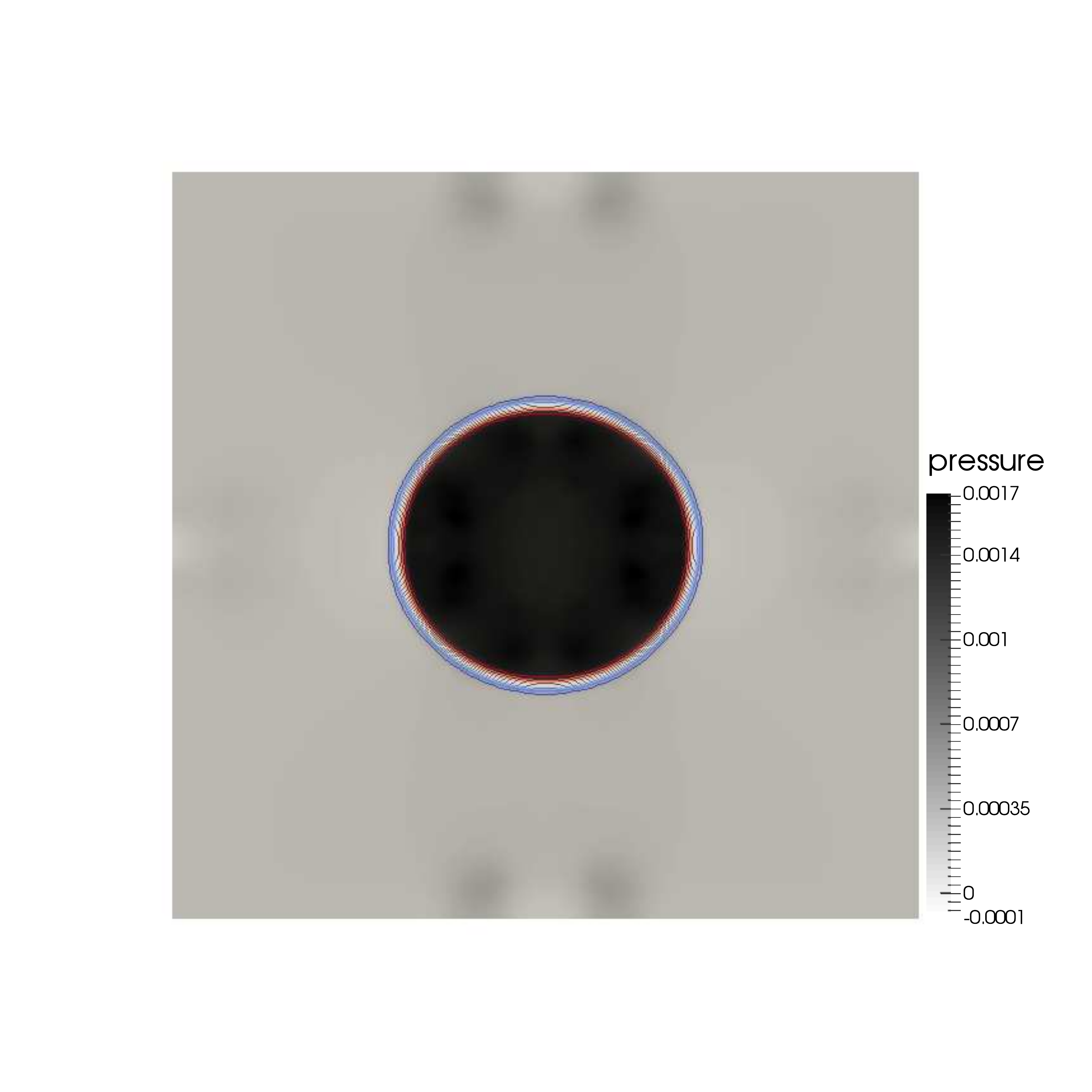}
  \caption{Pressure field and interface location at $t = 5T$ for the oscillating 
ellipse in a $128\times128$ mesh advected with a single recompression step.  
Left figure uses the standard 2nd order shift operator while the right one uses 
the newly developed energy-preserving one. Contour lines are plotted for 
$\Delta\ls = 0.1$.}
  \label{fig:ellipse-fields-1-128-41}
\end{figure}

The results obtained by including a single recompression step into the algorithm 
are presented in Figure~\ref{fig:EnergyEvolution-ellipse-rec}. They show how, 
irrespective of the use of an energy-preserving scheme into the advection 
scheme, the amount of energy included into the system in order to keep a sharper 
profile results in a small, but non-physical, increase of mechanical energy.  
Compared with Figure~\ref{fig:EnergyEvolution-ellipse-adv}, it can be seen how 
the difference is not as much in mechanical energy but rather \added[id=R1]{in 
}the nature of the oscillations. While results without recompression still 
preserve to some extent the oscillating nature of the physical system, 
recompression \replaced[id=R1]{produces}{produce} an enhanced smoothing, 
resulting in a flat profile in terms of both kinetic and potential energy.

The claim stated above can be clearly seen in
Figure~\ref{fig:ellipse-fields-1-128-41}, where the initial ellipsoid, expected 
to present a dynamical equilibrium, results in a fully rounded shape.
Besides, Figure~\ref{fig:ellipse-fields-1-128-41} \replaced[id=R1]{shows}{show} 
how the resulting fields, in both cases, are irrespective of the interpolation 
scheme for curvature used for the advection scheme. Further discussion on the 
impact of recompression in the final result is discussed in Section 
\ref{sec:Conclusions}. 

\FloatBarrier
\fi
\ifCapillaryWave
\subsection{Capillary wave}
A pure capillary wave is set by originally locating the interface at
$x = 0.2sin(ky)$, producing an initial wave along the vertical center line of 
wavelength $2\pi/k$. We set $k=\pi/H$, so that a single oscillating period is 
contained within the domain. Velocity is initially at rest.  With the mentioned 
boundary and initial conditions, the wave is expected to oscillate indefinitely, 
alternating states of maximum potential energy (i.e., maximum elongation) and 
minimum kinetic energy (i.e., fluid at rest) and vice-versa.  Initial setup is 
presented in Figure~\ref{fig:CapillaryWave-InitialSetup}.

\replaced[id=R1]{As is well known}{Well-known} from linear perturbation 
theory~\cite{Lamb1945}, the oscillation of the given setup present a 
characteristic period of $T=2\pi\sqrt{2\rho/\gamma k^3tanh(kH)}$, which is used 
as the reference value for time. On the other hand, the characteristic length 
scale is $L=2\pi/k$, the wavelength of the perturbation.  This yields a 
characteristic velocity of
$c=L/T= \sqrt{\gamma ~ k ~ tanh(kH)/2\rho}$, while pressure is referenced to 
$\rho c^2$, where $\rho$ stands for the average. Integration in time is set to 
$2T$.

\begin{figure}[h]
  \centering
  \includegraphics[width=5cm, trim={4cm 1cm 0cm 4cm},clip] {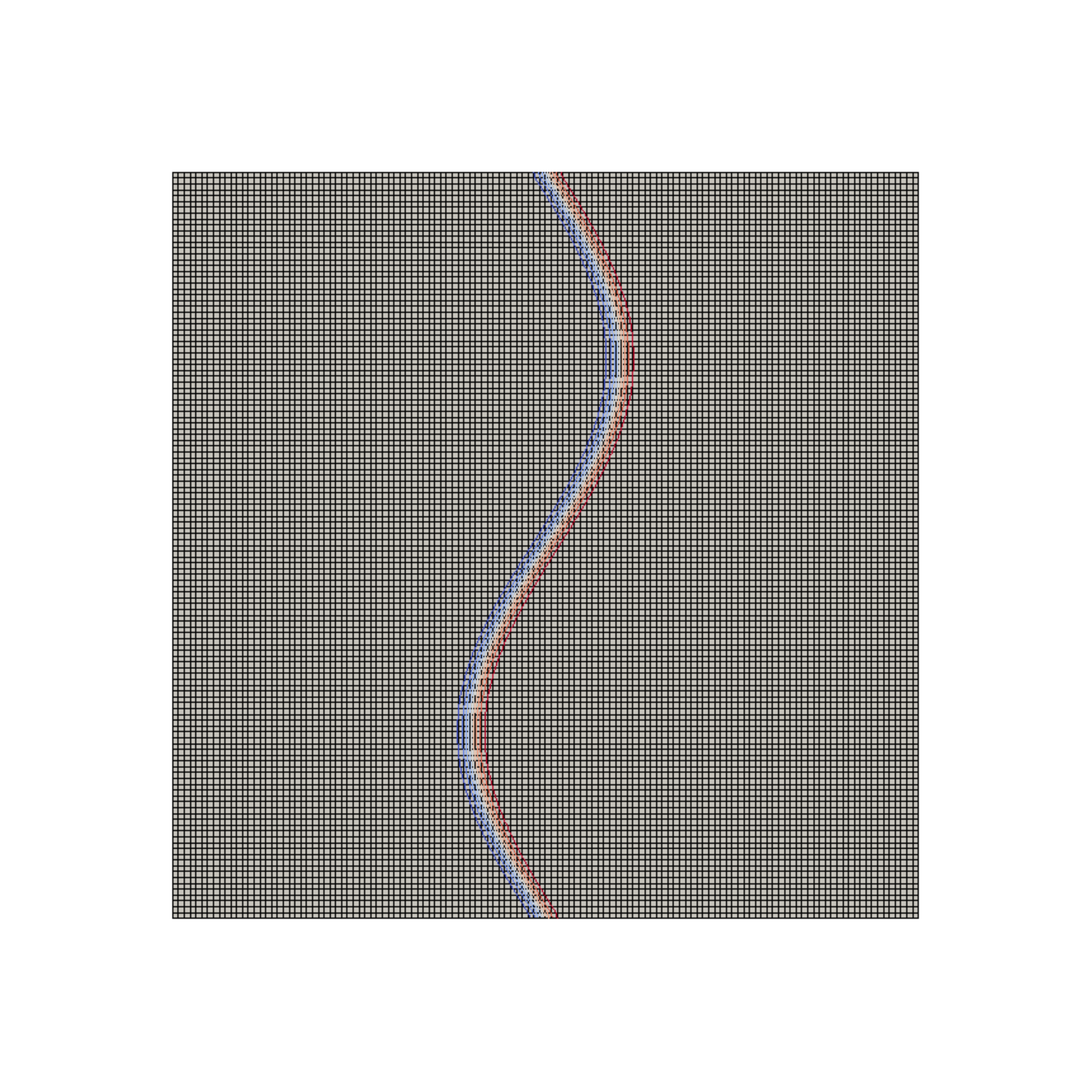}
  \caption{Initial setup of the marker function in a $128\times128$ mesh.   
 Contour lines are plotted for $\Delta\ls = 0.1$.}
  \label{fig:CapillaryWave-InitialSetup}
\end{figure}

Results in Figure~\ref{fig:EnergyEvolution-film-adv} show how the energy 
preserving discretization proposed in the present work preserves mechanical 
energy (top row, solid line) by balancing the resulting energy transfers (bottom 
row, solid line). While the standard midpoint interpolation of curvature results 
in a non-physical increase of mechanical energy, which ultimately leads to 
instabilities, the novel proposed method provides a stable discretization.

\begin{figure}[h]
  \centering
  \includegraphics[width=\textwidth]{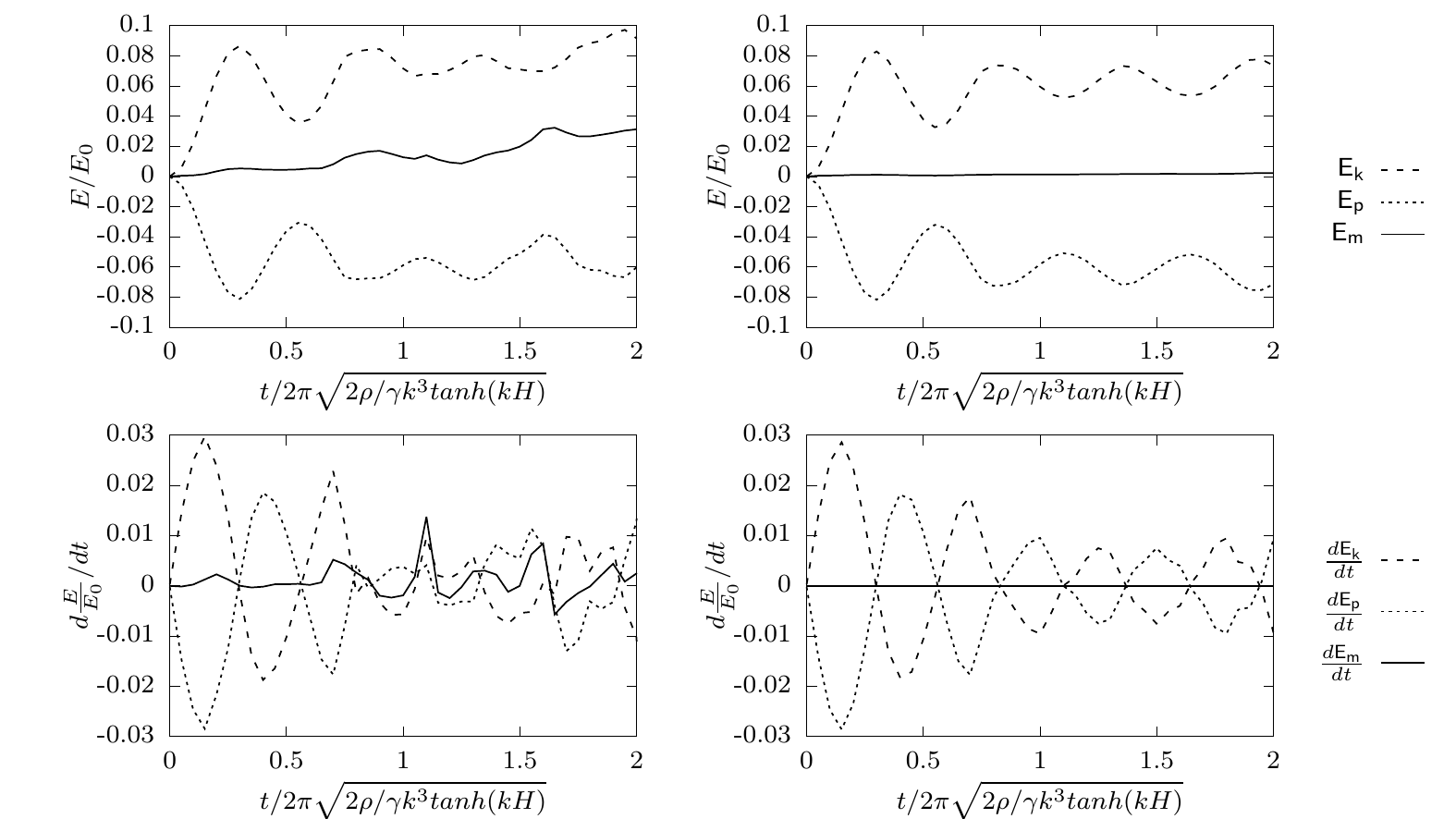}
  \caption{Energy evolution of the capillary wave for a pure advection case 
    (i.e., no recompression) with the standard interpolation of Olsson and 
    Kreiss \cite{Olsson2005} for the curvature (left) and the newly proposed 
    method (right) in a $128\times128$ mesh. Top rows show the discrete values 
    of kinetic ($\sfield{E_k}$), potential ($\sfield{E_p}$) and total 
    ($\sfield{E_m}$) energy.  Bottom rows show their semi-discretized time 
    derivative according to equations (\ref{eqn:dEkdt-multi-discrete}), 
  (\ref{eqn:dEpdt-multi-discrete}) and (\ref{eqn:dEmdt-multi-discrete}), 
respectively.}
  \label{fig:EnergyEvolution-film-adv}
\end{figure}
\begin{figure}[h]
  \centering
  \includegraphics[width=0.49\textwidth, trim={4cm 1cm 0cm 4cm} ,clip] 
  {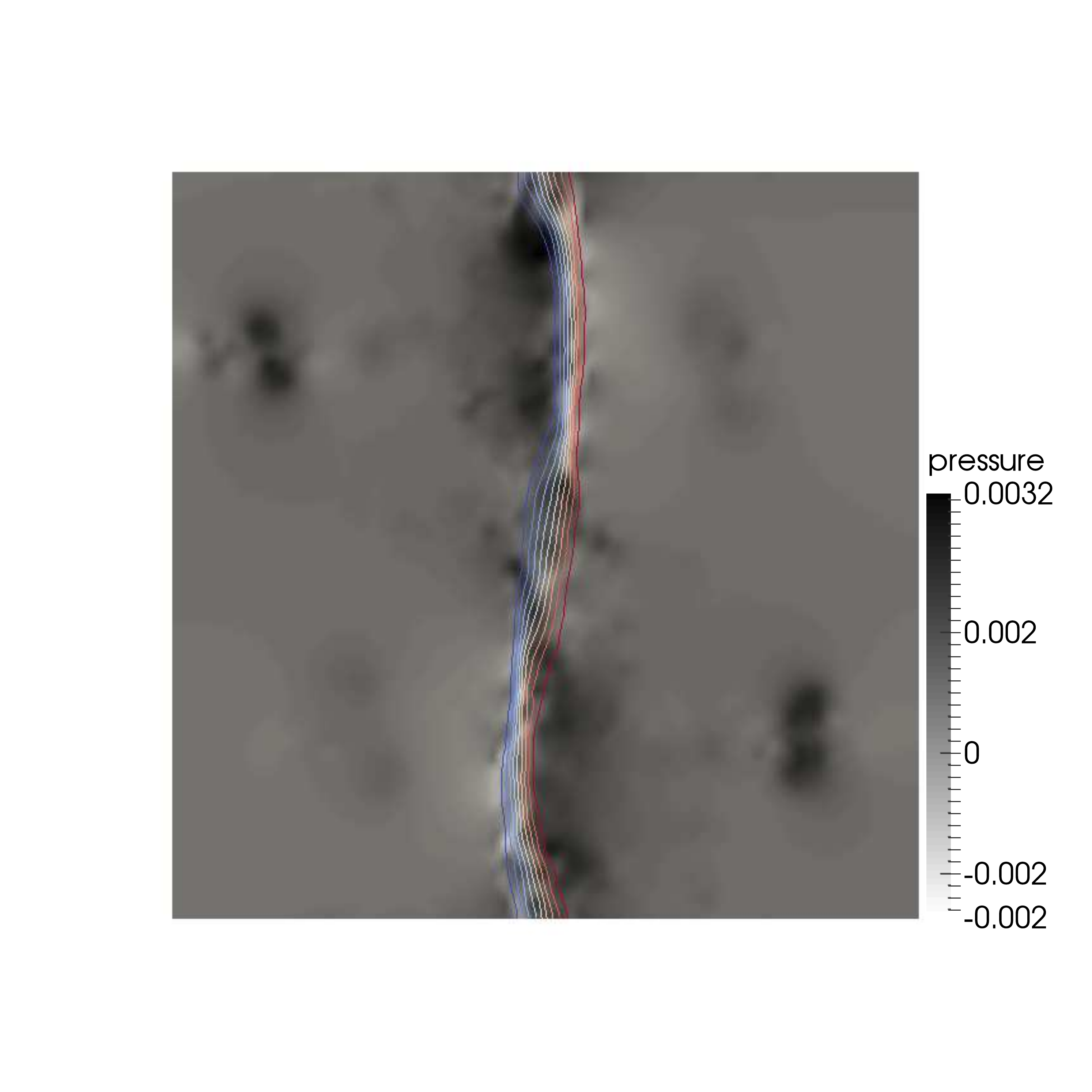}
  \includegraphics[width=0.49\textwidth, trim={4cm 1cm 0cm 4cm} ,clip] 
  {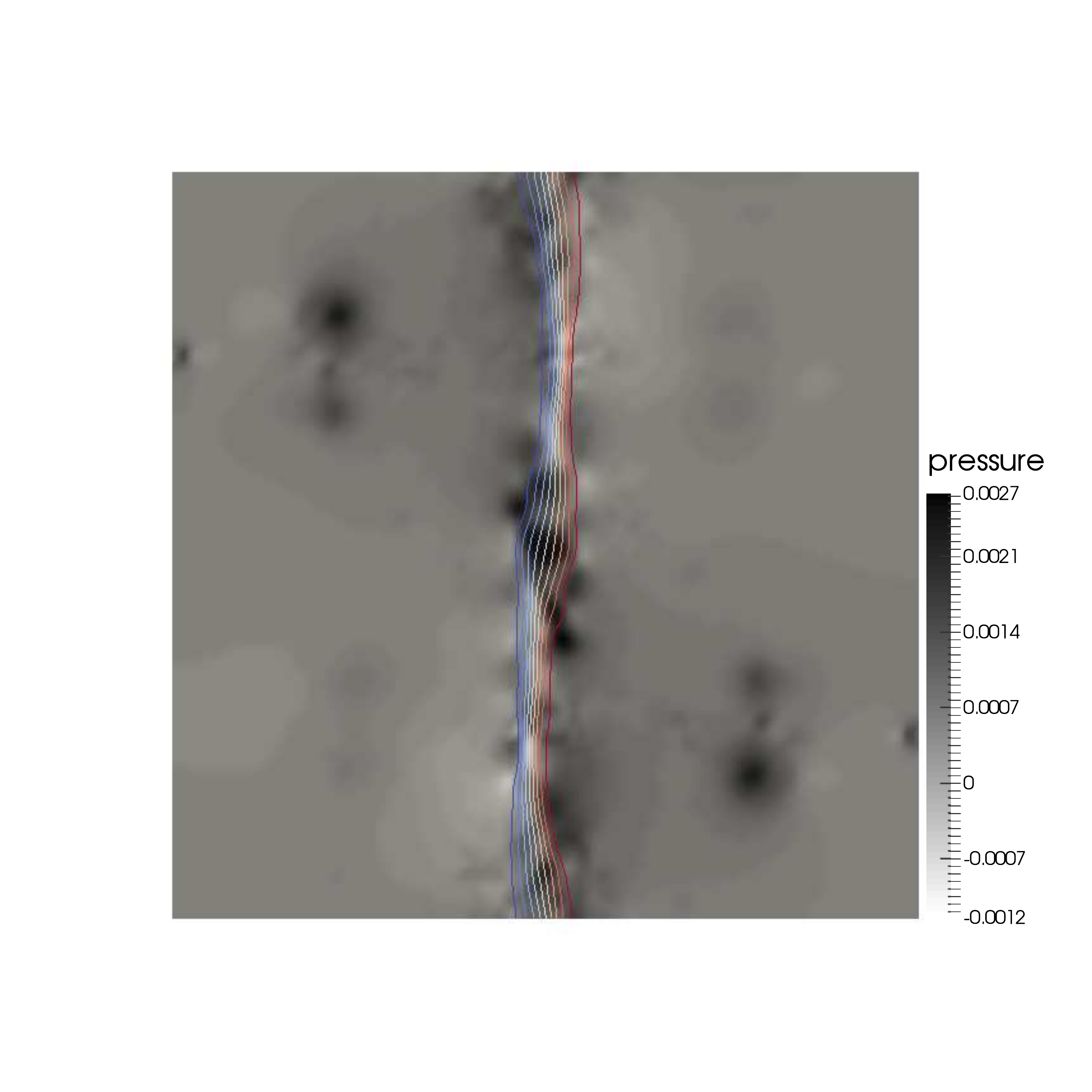}
  \caption{Pressure field and interface location at $t = 2T$ for a pure 
  capillary wave in a $128\times128$ mesh advected without recompression.  Left 
figure uses the standard 2nd order shift operator while the right one uses the 
newly developed energy-preserving one. Contour lines are plotted for $\Delta\ls 
= 0.1$.}
  \label{fig:film-fields-0-128-41}
\end{figure}

Even when mechanical energy is conserved in the newly proposed method, both the 
amplitude of kinetic and potential oscillations 
(Figure~\ref{fig:EnergyEvolution-film-adv}, top row, right) and the magnitude of 
the energy transfers (Figure~\ref{fig:EnergyEvolution-film-adv}, bottom row, 
right) exhibit a significant damping. The reason behind such a damping is the 
non-null contribution of surface tension to the momentum equation (the desired 
result for a closed surface) which produces an artificial acceleration of the 
fluid. The origin of such artificial acceleration lies in the discretization of 
curvature, particularly the computation of normals, which is at the origin of 
the errors that propagate to the momentum equation. This non-physical increase 
in kinetic energy \replaced[id=R1]{manifests}{manifest} itself as an increase of 
the base level of kinetic energy at
off-peaks, as can be seen in the top row of 
Figure~\ref{fig:EnergyEvolution-film-adv}. While naive interpolation techniques 
are unresponsive to such energy increments, the new energy-preserving method 
adjusts the transfers between kinetic and potential energies through surface 
tension to keep mechanical energy constant. As a result, the artificial and 
progressive increase in the kinetic energy level leaves no room to capillary 
oscillations, driving the system to a stagnant, but stable, situation.

Figure~\ref{fig:EnergyEvolution-film-rec}, on the other hand, includes a 
recompression step into the evolution of the wave. Results show clearly how, 
despite its \replaced[id=R1]{known}{know} advantages \cite{Olsson2005}, the 
resulting solution does not preserve energetic balances but rather increase 
total energy of the system, leading to eventual instabilities. It can be seen 
how the gain in sharpness introduced by recompression schemes is at the expenses 
of  destroying the advantages of the energy-preserving discretization. Results 
in Figure~\ref{fig:film-fields-1-128-41} can be compared with those of 
Figure~\ref{fig:film-fields-0-128-41}, which \replaced[id=R1]{shows}{show} how 
recompression \replaced[id=R1]{increases}{increase} the total energy of the 
system.  Namely, the scale in Figure~\ref{fig:film-fields-1-128-41} shows how 
velocity magnitudes are clearly higher regardless of the advective step is 
energy-preserving or not.  Among them, the energy-preserving scheme shows milder 
velocity fields.  This role of recompression is analyzed in 
Section~\ref{sec:Conclusions}.

\begin{figure}[h]
  \centering
  \includegraphics[width=\textwidth]{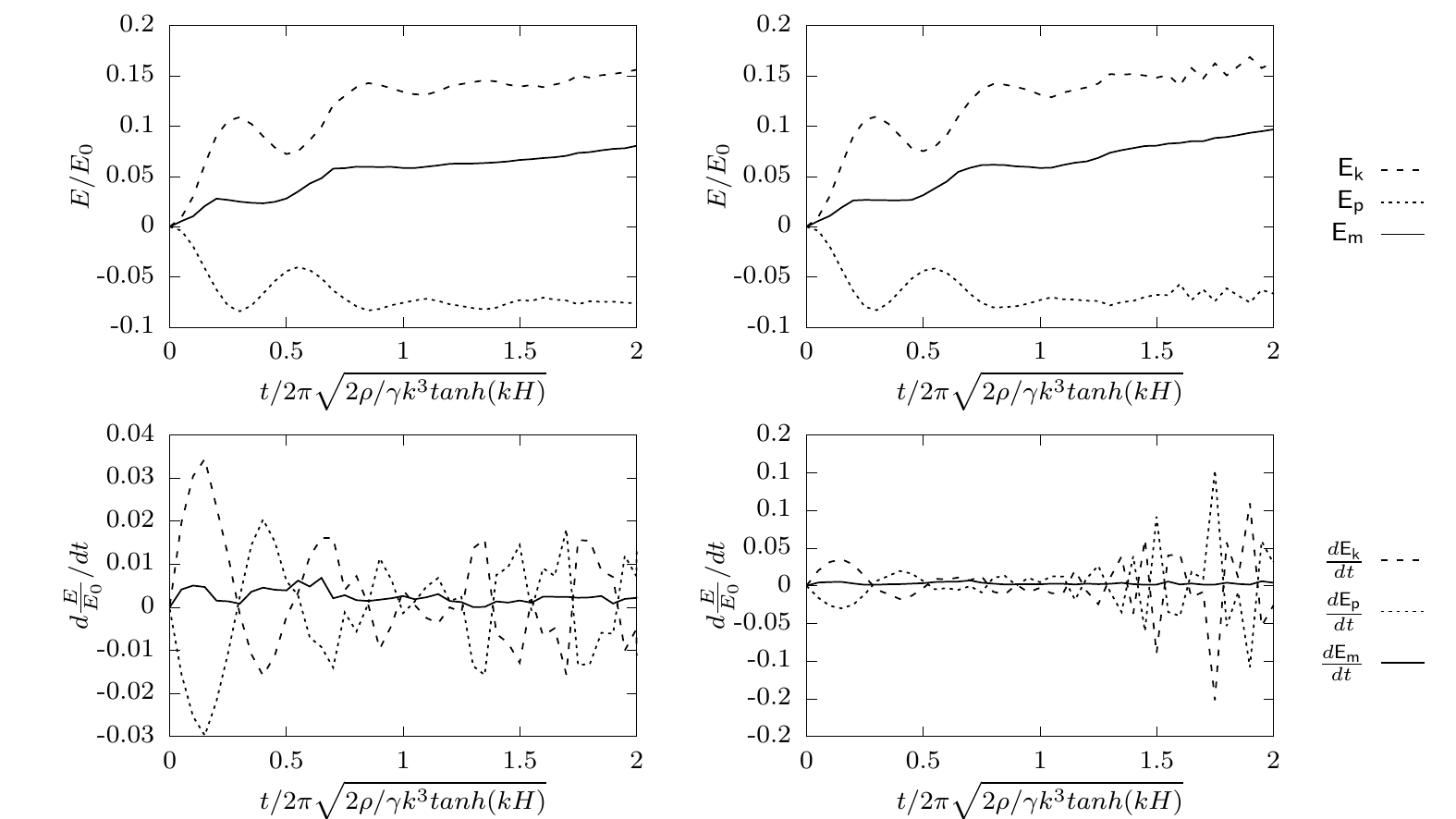}
  \caption{Energy evolution of the capillary wave with the complete Olsson and 
    Kreiss method \cite{Olsson2005} with a single recompression stage (left), 
    and the same method including the modified curvature interpolation (right) 
    in a $128\times128$ mesh. Top rows show the discrete values of kinetic 
    ($\sfield{E_k}$), potential ($\sfield{E_p}$) and total ($\sfield{E_m}$) 
    energy.  Bottom rows show their semi-discretized time derivative according 
    to equations (\ref{eqn:dEkdt-multi-discrete}), 
  (\ref{eqn:dEpdt-multi-discrete}) and (\ref{eqn:dEmdt-multi-discrete}), 
respectively.}
  \label{fig:EnergyEvolution-film-rec}
\end{figure}
\begin{figure}[h]
  \centering
  \includegraphics[width=0.49\textwidth, trim={4cm 1cm 0cm 4cm} ,clip] 
  {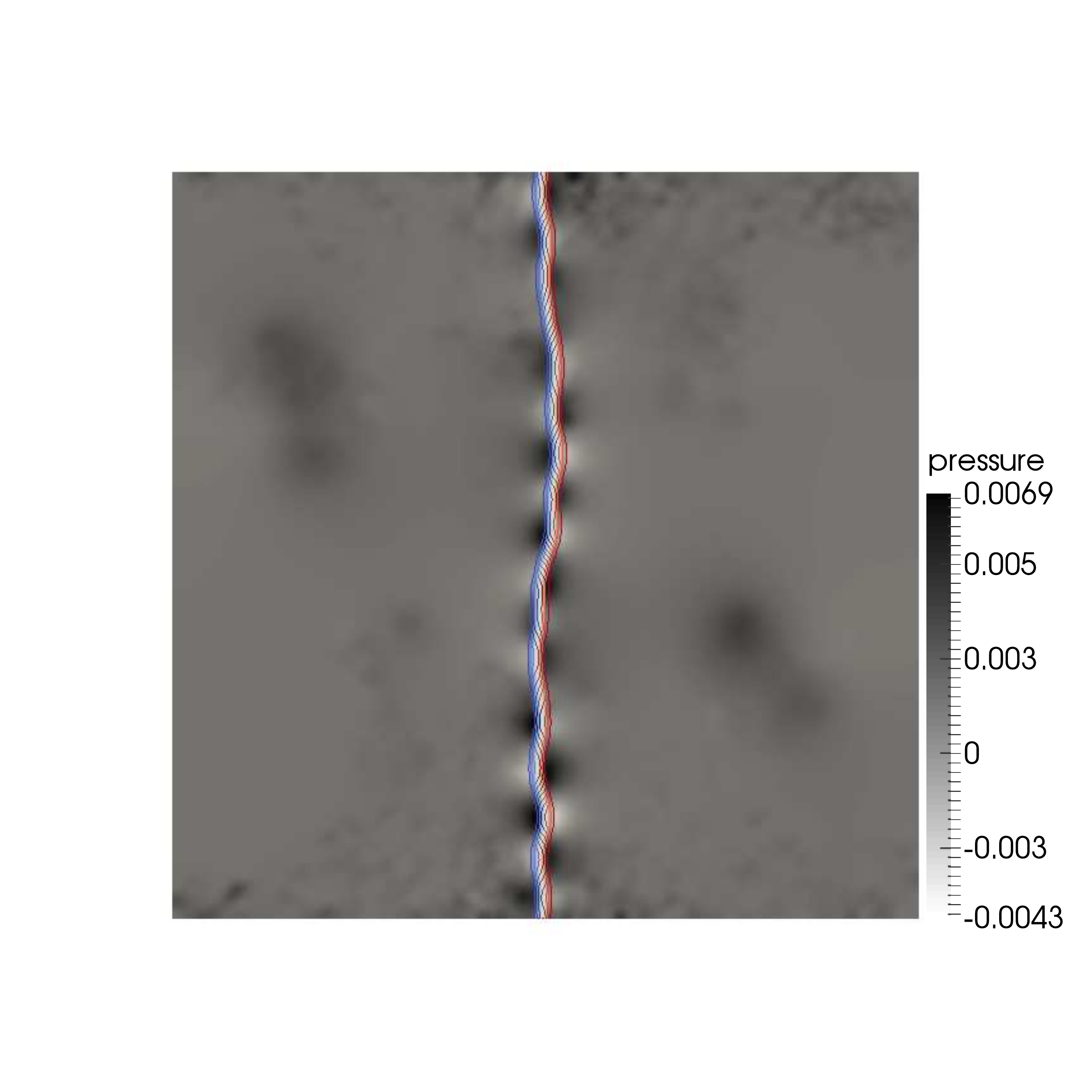}
  \includegraphics[width=0.49\textwidth, trim={4cm 1cm 0cm 4cm} ,clip] 
  {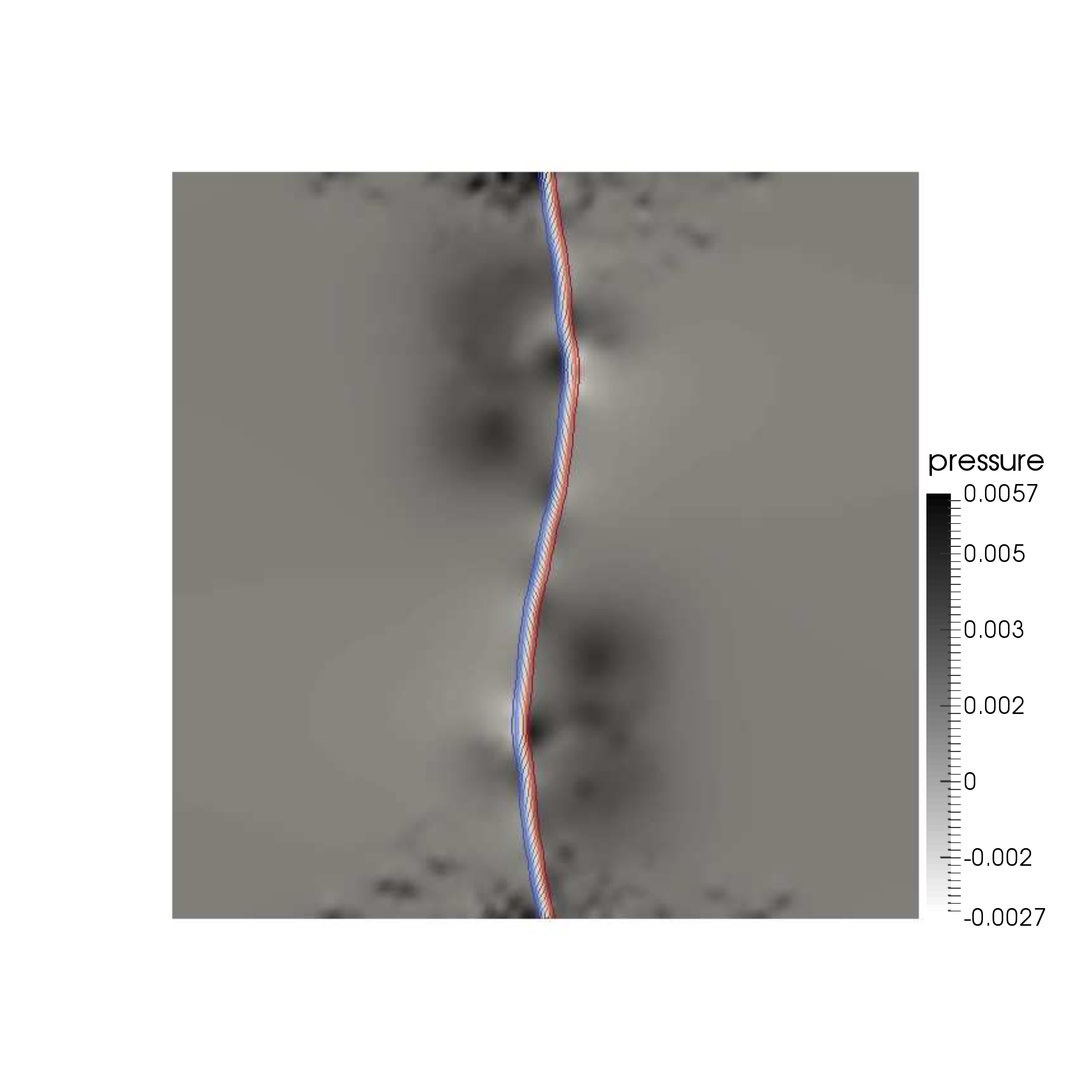}
  \caption{Pressure field and interface location at $t = 2T$ for a capillary 
  wave in a $128\times128$ mesh advected with a single recompression step.  Left 
  figure uses the standard 2nd order shift operator while the right one uses the 
  newly developed energy-preserving one. Contour lines are plotted for 
$\Delta\ls = 0.1$.}
  \label{fig:film-fields-1-128-41}
\end{figure}
\FloatBarrier
\fi

\section{Concluding Remarks}
\label{sec:Conclusions}

By incorporating the first variation of area, equation (\ref{eqn:dSdt}), into 
the continuum formulation we have explicitly imposed a novel condition to the 
system. Equation (\ref{eqn:dSdt-identity}) shows that the use of a smooth marker 
function is compatible with such a condition. This condition is implicitly 
incorporated into the discretized system by means of a newly developed curvature 
cell-to-face shift operator, $\E$, defined in equation 
(\ref{eqn:CurvatureInterpolator}).  Analytical and numerical assessments provide 
evidence that\added[id=R3]{, in the absence of recompression,} the novel 
interpolation scheme \replaced[id=A]{preserves}{preserve} mechanical energy up 
to temporal integration by balancing kinetic and potential energy transfers to 
machine accuracy.

The exact value of both kinetic and potential energy is not achieved due to the 
lack of conservation of linear momentum\replaced[id=A]{. This implies that,
while the transfers between surface and kinetic energy are equal and of opposite
sign, its magnitude is not necessarily the correct one.}{in the former,
which is a well-known issue in multiphase flows and the object of ongoing 
research \mbox{\cite{Kim2005, Abu-Al-Saud2018}}}.

\added[id=R1, remark={Partially rephrased from the original manuscript}]{
  In this regard, the adoption of a fully conservative momentum formulation, 
along with proper discretization techniques for the convective operator, as 
already announced in Section~\ref{sec:DiffModel_multi_EP}, should be considered 
in a general case.
 However, the formulation of the surface tension is the most challenging
term. Not being cast into a conservative form, it relies on the accurate
capturing of the interface to produce a closed, and thus conservative, force 
field.
 In summary, the use of a finite grid prevents us from resolving the finest 
scales of the interface, represented by the marker function $\ls$. This 
under-resolution of $\ls$, either induced both by the mesh and the advection 
scheme, induces subsequent errors in the computation of both $\ni$ and $\ki$, as
stated by Magnini et al.  \cite{Magnini2016}. These errors spread  into the 
momentum equation, which can be seen as a back-scatter of energy from the 
finest, unresolved, surface representation scales into larger kinematic ones, 
manifesting itself as an inappropriate momentum balance, which ultimately leads 
to an inaccurate kinetic energy level.
This is a well-known issue in multiphase flows and the object of ongoing 
research \cite{Kim2005, Abu-Al-Saud2018}.}

\added[id=R1]{
  Nevertheless, despite the lack of linear momentum conservation, mechanical 
energy is conserved and thus the stability of the system is guaranteed up to 
temporal integration.}
\added[id=R1]{From this perspective, the novel technique may provide extra 
reliability for surface energy governed phenomena, particularly those involving 
surface break-up or coalesce, as it may occur in atomization processes or 
Plateau-Rayleigh instabilities, among others.}

\replaced[id=R3]{
  Recompression schemes, despite producing an energetic imbalance, as has been 
 shown in Section~\ref{sec:Results}, are common in the level-set community. They
 preserve a coherent marker field at the expenses of introducing non-physical 
 energy to the system. Even when the proposed method enforces the energetic 
 consistency between marker and momentum transport equations, the inclusion of 
 recompression prevents us from obtaining a fully energy-preserving scheme.  }
{The role of interface recompression deserves a special remark.  It has been 
  included in the level set literature \mbox{\cite{Sussman1994, Olsson2005}} in 
  order to remedy the distortion that
  the convective schemes may induce. Even when conserving mass, such as in 
  \mbox{\cite{Olsson2005, Olsson2007}}, this results in a non-null contribution
  to potential energy. Actually, Section~\ref{sec:Results} shows this artificial
  contribution to the total energy of the system.}
Following the spirit described in Section~\ref{sec:DiffModel_multi}, a first 
approach may be to modify the recompression step to produce not only a 
conservative, but an energetically neutral resharpening.
Enforcing a null contribution to potential energy of equation 
(\ref{eqn:CLS-recompression}),
if possible, would allow an arbitrary number of recompression steps, avoiding 
any penalty in terms of energetic balances. Although this would be desirable, it 
requires to re-formulate a mass- and energy- conservative recompression scheme 
which effectively moves the interface irrespective of advection, which is 
definitively not obvious.

Others have tried to include recompression within the advection step to yield a 
single-step method. After all, recompression is included to fix the 
\added[id=A]{distortion produced by }interface advection. This leads to 
phase-field-like methods \cite{Guermond2017, Mirjalili2018}. Interpreting this 
idea as a custom-made high resolution scheme, these approaches 
\replaced[id=A]{can}{may} eventually be cast into a convective form like that in 
equations~(\ref{eqn:ConvectionSplit}) and proceed to obtain the equivalent 
curvature interpolation as in equation~(\ref{eqn:CurvatureInterpolator}). A 
variant of this model may be to approach the advection of the marker function as 
a regularization problem \cite{Trias2018}.

\deleted[id=R1]{In summary, the use of a finite grid prevents us from resolving 
the finest scales of the interface, represented by the marker function $\ls$.  
This under-resolution of $\ls$, either induced by the mesh or the advection 
scheme, induces subsequent errors in the computation of both $\ni$ and $\ki$, as 
stated by Magnini et al.  \cite{Magnini2016}. These errors spread  into the 
momentum equation, which can be seen as a back-scatter of energy from the 
finest, unresolved, surface representation scales into larger kinematic ones, 
manifesting itself as an inappropriate momentum balance, ultimately leading to 
an inaccurate kinetic energy level.
In this regard, recompression schemes, in its ambition to preserve a coherent 
marker field, do so at the expenses of introducing non-physical energy to the 
system, resulting in the announced increase in mechanical energy.}


Lastly, both a review of the well-known symmetry-preserving scheme and the 
development of the energy-preserving scheme have been approached from an 
algebraic point \added[id=A]{of }view. Aside from the advantages in terms of 
algebraic analysis, \deleted[id=A]{from a purely computational perspective }the 
use of an \replaced[id=A]{algebra-}{algebraic-}based discretization provides an 
opportunity for High Performance Computing (HPC) optimization, parallelization 
and portability \cite{Alvarez2018}. By casting differential forms into algebraic 
ones, (i.e., matrices and vectors), it has been shown in \cite{Alvarez2018} that 
nearly 90\% of the operations comprised in a typical FSM algorithm for the 
solution of incompressible Navier-Stokes equations can be reduced to 
\deleted[id=A]{the following: }Sparse Matrix-Vector multiplication (SpMV), 
generalized vector addition (AXPY) and dot product (DOT).
In this regard, the present formulation falls within a smart strategy towards 
portable, heterogeneous, HPC.

%

%


\section{Acknowledgements}
\label{sec:Acknowledgments}

This work has been financially supported by the Ministerio de Economía y 
Competitividad, Spain (ENE2017-88697-R and ENE2015-70672-P) as well as an FI 
AGAUR-Generalitat de Catalunya fellowship (2017FI\_B\_00616) and a Ramón y Cajal 
postdoctoral contract (RYC-2012-11996). The authors are grateful to Prof. Roel 
Verstappen, Prof. Arthur Veldman and MSc Ronald Remmerswaal for their enriching 
discussions.

\appendix
\newcommand{\ifAlgebraicAnalysisL}{\iffalse}
\section{Inner products}
\label{sec:InnerProducts}

Inner products are bilinear maps from a vector space to its base field (i.e.,  
$\DOT[S]{\cdot}{\cdot}: \set{S} \times \set{S} \to \Field$). Inner products can 
be defined over both continuum and discrete spaces as
\begin{equation}
  \DOT[S]{f}{g} = \int_{\set{S}} fg d\set{S} \quad \forall f,g \in \set{S}
  \label{eqn:InnerProduct-definition}
\end{equation}

This  definition can be readily applied to discrete fields, yielding the 
 definition of inner products for discrete vectors within metric spaces as
\begin{equation}
  \DOT[S]{\sfield[s]{f}}{\sfield[s]{g}} = {\sfield[s]{f}}^T \metric{S} 
  \sfield[s]{g}
  \label{eqn:InnerProduct-definition-discrete}
\end{equation}
where $\metric{S}$ takes over the role of integrating in space, whereas the 
transpose of the first element provides with the appropriate order to perform 
the subsequent products and sums. This can be seen by expressing $f$ and $g$ as 
a finite sum of piecewise defined base functions.

\added[id=R1]{Within this framework, we can define skew-symmetry as the property
  of operators satisfying}
  {
    \begin{equation}
      \DOT{\phi}{A\psi} = -\DOT{A\phi}{\psi} \quad \forall \phi,\psi \in \set{S} 
      \quad A:\set{S} \to \set{S}
      \label{eqn:Skew-symmetry-defintion}
    \end{equation}
  }
\added[id=R1]{where, in the discrete setting, $A$ must be a skew-symmetric 
matrix. Similarly, we can define duality as}
{
    \begin{equation}
      \DOT{\phi}{A\psi} = \DOT{A^*\phi}{\psi}
      \quad \forall \phi \in \set{S} \psi \in \set{T}
      \quad A:  \set{T} \to \set{S}
      \quad A^*:\set{S} \to \set{T}
      \label{eqn:Duality-defintion}
  \end{equation}}

\deleted[id=A]{Recovering the initial definition it can be analyzed what is 
their role in duality relations. }By using the aforementioned 
\replaced[id=A]{definitions}{definition of inner product} and the well-known 
Gauss-Ostrogradsky theorem, it provides with
\begin{equation}
  \DOT{f}{\nabla \cdot \vec{g}}
      = \int_\Omega f \nabla \cdot \vec{g}
      = - \int_\Omega \nabla f \cdot \vec{g}
        + \int_{\partial \Omega} f \vec{g} \cdot \hat{n}
      \label{eqn:InnerProduct-duality}
\end{equation}
where we can see that, assuming that there are no contributions from the 
boundary, the usual relation $\DOT{f}{\nabla \cdot \vec{g}} = -\DOT{\nabla 
f}{\vec{g}}$ holds as usual. However, if there is a discontinuity in either $f$ 
or $\vec{g}$ as a consequence of, say, an interface ($\Gamma$) in the domain, 
this prevents us from using equation (\ref{eqn:InnerProduct-duality}) directly, 
but rather first in both sub-domains separately and then sum them together. This 
results in an explicit expression as
\begin{equation}
  \DOT{u}{\nabla \cdot \vec{v}}
      = \int_\Omega             u \nabla \cdot \vec{v}
      = \int_{\partial \Omega}  u \vec{v} \cdot \hat{n}
      + \int_{\Gamma}           \left[ u \vec{v} \right] \cdot \hat{n}
      - \int_\Omega             \nabla u \cdot \vec{v}
  \label{eqn:InnerProduct-duality-discontinuity}
\end{equation}
where the discontinuity is now explicitly included in the system. Note then that 
for a discrete system, the aforementioned gradient-divergence duality is
\begin{equation}
  \DOT[S]{\sfield[s]{u}}{\DIV\vfield[s]{v}}
  = - \DOT[S]{\GRAD \sfield[s]{u}}{\vfield[s]{v}}
    + \int_\Gamma {\left[ \sfield[s]{u}\vfield[s]{v} \right]}
  \label{eqn:Duality-discrete-discontinuous}
\end{equation}
where the extra rightmost term captures the corresponding jump of the variables 
under consideration. Note that a proper approximation of $\Gamma$ is required in 
order to obtain accurate solutions.

\ifAlgebraicAnalysisL
\section{Algebraic Analysis of L}
\label{sec:Differential-Algebra}
Mimicking the properties of the continuum operator, $\LAP$ should be 
self-adjoint, i.e.:

\begin{equation}
  \DOT[C]{\LAP \sfield[c]{u}}{\sfield[c]{v}} = \DOT[C]{\sfield[c]{u}}{\LAP 
  \sfield[c]{v}}
  \label{eqn:Self-AdjointnessCondition}
\end{equation}

We can check that this holds if $\metric{C}\LAP$ (i.e., the integrated 
discretization) is symmetric ($\metric{C}\LAP = \LAP^T \metric{C}$). Remember 
that by definition $\metric{C}$ is a symmetric and positive-definite matrix.  
This is guaranteed by construction by means of equations
(\ref{eqn:DiscreteGradient}) and (\ref{eqn:DiscreteDivergence}), as far as 
$\Ci{F}$, $\Sf$ and $\Df$ are symmetric (in this case, diagonal) matrices.  
Indeed, this implies

\begin{equation}
  img(L) \perp ker(L)
  \label{eqn:ImgKernOrthogonal}
\end{equation}

In addition, because $\LAP = -\metric{C}^{-1} \topo{FC} \Sf (\Df)^{-1} 
\topo{CF}$, recalling that $\topo{FC} = \topo{CF}^T$ and that $\Sf$ and $\Df$ 
are diagonal and positive-definite, we can state that the Laplacian operator is 
a negative semi-definite operator. 

Next is to check the rank of the operator. For doing this, we proceed as  
follow:

\begin{equation}
  \LAP \sfield[c]{u} = \Zv{C} \quad \forall \sfield[c]{u} \in ker(\LAP)
  \label{eqn:KernelDefiniton}
\end{equation}

Then,
\begin{equation}
  \DOT[C]{\sfield[c]{u}}{\LAP \sfield[c]{u}} = 0 \quad \forall \sfield[c]{u} \in 
  ker(\LAP)
  \label{eqn:KernelLaplacianCondition}
\end{equation}

Which can be decomposed as:

\begin{equation}
  -\left(\topo{CF} \sfield[c]{u} \right)^T \Sf (\Df)^{-1} \left(\topo{CF} 
   \sfield[c]{u} \right) = 0
  \label{eqn:LaplacianDecompositionBilinearProduct}
\end{equation}

This turns into:
\begin{equation}
  \sum_{f} \dvec[f]{s} \dvec[f]{d}^{-1} \left( u_+ - u_- \right)^2 = 0
  \label{eqn:LaplacianDecompositionBilinearProductFinal}
\end{equation}

Where it can be seen that $u_+ = u_-$ and so $\uc$ must be a multiple of 
$\Iv{C}$.  Note that this result holds if the domain is connected, a well-known 
result for Laplacian matrices. Accordingly, it can be stated that:

\begin{equation}
  null(\LAP) = 1
  \label{eqn:LaplacianNullity}
\end{equation}

Which, used together with equation (\ref{eqn:ImgKernOrthogonal}) means that 
\emph{any vector orthogonal to $\Iv{C}$ belongs to $img(\LAP)$}. This can be 
used to state solvability conditions. In a general system $\LAP \uc = 
\sfield[c]{b}$, $\sfield[c]{b} \in img(\LAP)$ to have a solution. This can be 
easily verified by checking $\DOT{\Iv{C}}{\sfield[c]{b}}$.
\fi

\bibliographystyle{model1-num-names}
\bibliography{library}
\end{document}